\documentclass[%
 aip,
 jcp,%
 amsmath,amssymb,
reprint,%
longbibliography
]{revtex4-1}

\usepackage{graphicx}% Include figure files
\usepackage{epstopdf}
\usepackage{dcolumn}% Align table columns on decimal point
\usepackage{bm}% bold math
\usepackage{hyperref}% add hypertext capabilities
\usepackage{comment}
\usepackage{color}
\newcommand{\im}{\mathrm{i}}
\newcommand{\dif}{\mathrm{d}}

\newcommand{\e}{\mathrm{e}}

\begin{document}

\title{Exact description of excitonic dynamics in molecular aggregates weakly driven by light}

\author{Veljko Jankovi\'{c}}
\email{veljko.jankovic@ipb.ac.rs}
\affiliation{
 Faculty of Mathematics and Physics, Charles University, Ke Karlovu 5, 121 16 Prague 2, Czech Republic
}
\affiliation{Scientific Computing Laboratory, Center for the Study of Complex Systems, Institute of Physics Belgrade, University of Belgrade, Pregrevica 118, 11080 Belgrade, Serbia}
\author{Tom\'{a}\v{s} Man\v{c}al}%
 \email{mancal@karlov.mff.cuni.cz}
\affiliation{
 Faculty of Mathematics and Physics, Charles University, Ke Karlovu 5, 121 16 Prague 2, Czech Republic
}

\begin{abstract}
We present a rigorous theoretical description of excitonic dynamics in molecular light-harvesting aggregates photoexcited by weak-intensity radiation of arbitrary properties. While the interaction with light is included up to the second order, the treatment of the excitation--environment coupling is exact and results in an exact expression for the reduced excitonic density matrix that is manifestly related to the spectroscopic picture of the photoexcitation process. This expression takes fully into account the environmental reorganization processes triggered by the two interactions with light. This is particularly important for slow environments and/or strong excitation--environment coupling. Within the exponential decomposition scheme, we demonstrate how our result can be recast as the hierarchy of equations of motion (HEOM) that explicitly and consistently includes the photoexcitation step. We analytically describe the environmental reorganization dynamics triggered by a delta-like excitation of a single chromophore, and demonstrate how our HEOM, in appropriate limits, reduces to the Redfield equations comprising a pulsed photoexcitation and the nonequilibrium F\"{o}rster theory. We also discuss the relation of our formalism to the combined Born--Markov--HEOM approaches in the case of excitation by thermal light.
\end{abstract}

%\keywords{Suggested keywords}%Use showkeys class option if keyword
                              %display desired
\maketitle

\section{\label{sec:Intro} Introduction}
Recent years have seen vigorous interest in unveiling the basic physical mechanisms governing the electronic solar energy conversion in photosynthetic systems.~\cite{Quantum-Effects-in-Biology,RevModPhys.90.035003,AnnuRevPhysChem.66.69,JRSocInterface.11.20130901} The developments in this field are expected to provide new ways of improving the light-to-charge conversion in artificial systems, e.g., organic photovoltaics (OPVs).~\cite{NatMater.16.35}
A thorough understanding of the solar energy
conversion in molecular light-harvesting systems calls for a detailed description of light absorption,
excitation energy transfer (EET), charge separation, and charge transport.~\cite{Blankenship-book,AccChemRes.42.1691} Our current understanding of these steps
has been shaped by ultrafast spectroscopy experiments, which can provide insights into the dynamics of electronic excitations on time scales as short as a couple of femtoseconds.~\cite{Nature.446.782,NatChem.6.706,Science.323.369,Science.344.1001} Such experiments, therefore, can also temporally resolve nuclear motions induced by photoexcitation, i.e., nuclear reorganization processes, which take place on $\sim$10--100 fs time scales. Moreover, photosynthetic EET falls into the so-called intermediate coupling regime,~\cite{RevModPhys.90.035003,AnnuRevPhysChem.66.69,AnnuRevCondensMatterPhys.3.333} in which the energy scales representative of electronic couplings, excitation--environment couplings, and static disorder in local transition energies are comparable to one another. Correspondingly, a proper interpretation of ultrafast experimental signatures necessitates development of explicitly time-dependent theoretical approaches that can accurately capture the non-Markovian dynamical interplay between temporal evolution of electronic excitations and their environment.~\cite{AnnuRevCondensMatterPhys.3.333,AnnuRevPhysChem.66.69} Examples of such methods include hierarchical equations of motion (HEOM),~\cite{JPhysSocJpn.75.082001}
and some wavefunction-based methods.~\cite{PhysRep.324.105} 

Apart from the nonperturbative treatment of the interaction with the environment, a comprehensive theoretical analysis of the dynamics of electronic excitations created during ultrafast spectroscopy experiments should explicitly consider the exciting radiation field. However, the explicit inclusion of the photoexcitation process has received only a limited attention so far. The photoexcitation is commonly assumed to be infinitely short, i.e., delta-like, so that it instantaneously produces excited-state populations, whose further evolution on ultrashort time scales is followed.~\cite{ProcNatlAcadSci.106.17255,PhysRevLett.121.026001} On the other hand, theoretical methods of nonlinear spectroscopy,~\cite{Mukamel-book} which explicitly keep track of the interaction with exciting pulses, have been employed in conjunction with, e.g., HEOM, to examine certain features of spectroscopic signals.~\cite{JChemPhys.134.194508} However, there has not been much discussion on how to explicitly include the photoexcitation and respect the nonperturbative treatment of the excitation--environment coupling.~\cite{PhysRevLett.123.093601}
The importance of the photoexcitation step is typically discussed within the debate on the relevance of the results of ultrafast experiments for the photosynthetic operation \emph{in vivo}.~\cite{NewJPhys.12.065044,ChemPhys.439.100,ProcNatlAcadSci.109.19575,JPhysChemLett.4.362,JPhysChemLett.9.2946} It is argued that, due to different properties of natural Sunlight compared to laser pulses employed in experiments, photoexcitation of photosynthetic complexes under natural conditions triggers different dynamics from the one observed in ultrafast experiments. Nevertheless, under the common assumption that the electronic system is initially unexcited, any nontrivial dynamics under both excitation conditions is ultimately induced by the interaction with the radiation. In a nonlinear spectroscopy experiment, the signal depends on the appropriate power of the exciting field, i.e., the perturbation expansion in the interaction with radiation is appropriate.~\cite{Mukamel-book,Mancal-chapter-2014} Similarly, the weakness of the excitation of photosynthetic complexes under natural conditions makes the second-order treatment of the interaction with light plausible.~\cite{NewJPhys.12.065044,ChemPhys.439.100,ProcNatlAcadSci.109.19575,JPhysChemLett.9.2946}

Indeed, it has been shown~\cite{NewJPhys.12.065044} that the excited-state dynamics of a molecular system weakly driven by light of arbitrary properties is completely determined by the first-order radiation correlation function and the reduced evolution superoperator. It can be said that the information required for constructing the dynamics under arbitrary (weak) driving can only be obtained by ultrafast spectroscopy,~\cite{ChemPhys.439.100} which provides access to the reduced evolution superoperator.
However, the analysis conducted in Ref.~\onlinecite{NewJPhys.12.065044} is quite general and does not provide any details on the form and properties of this superoperator. Certainly, it should contain information about the nonequilibrium evolution of the environment taking place between consecutive interactions with light.~\cite{ChemPhysLett.530.140} Along these lines, attempts have been made to examine the importance of these dynamical environmental effects for the second-order light-induced dynamics by augmenting the usual quantum master equation by terms that depend on the delay between the two interactions.~\cite{ChemPhys.404.103} The analysis of the second-order photoinduced dynamics in Ref.~\onlinecite{SciRep.6.26230} suggested that the nonequilibrium bath evolution between the two interactions with light is reflected in the so-called photoinduced correlation term. Let us note that the analyses conducted in Refs.~\onlinecite{ChemPhysLett.530.140,ChemPhys.404.103,SciRep.6.26230} are essentially perturbative in the excitation--environment coupling.

On the other hand, in the field of ultrafast semiconductor optics,~\cite{RevModPhys.70.145,RevModPhys.74.895,RepProgPhys.67.433} the photoexcitation step and the nonequilibrium dynamics of thus induced electronic excitations are typically studied within the density matrix (DM) theory complemented with the so-called dynamics controlled truncation (DCT) scheme.~\cite{RevModPhys.70.145,PhysRevB.53.7244} The DCT scheme classifies DMs according to the lowest power with which they scale in the exciting field and, therefore, provides a recipe to analyze the dynamics up to any given order in the exciting field in terms of a finite number of electronic DMs. A DCT-based approach has been recently applied by one of us to study exciton generation and subsequent charge separation in photoexcited OPVs.~\cite{PhysRevB.95.075308,JPhysChemC.121.19602} However, the truncation of the environment-assisted branch of the hierarchy within the DCT scheme still has to be performed separately,~\cite{RepProgPhys.67.433} and it is commonly done in a low order in the excitation--environment coupling.~\cite{PhysRevB.95.075308,PhysRevB.65.035303,RevModPhys.70.145} In a similar vein, the explicit consideration of the excitation by incoherent light is also combined with a Redfield-like treatment of the excitation--environment coupling.~\cite{JChemPhys.152.154101,JChemPhys.148.124114,JPhysB.50.184003} The individual excitation and deexcitation events may be treated within the Born--Markov quantum optical approximation~\cite{Breuer-Petruccione-book,PhysRevLett.113.113601} or by constructing the Bloch--Redfield quantum master equation.~\cite{JChemPhys.148.124114,JChemPhys.142.104107} Adopting the standpoint of the theory of open quantum systems, the effects of incoherent radiation may be taken into account by introducing an appropriate spectral density of the light--matter coupling.~\cite{PhysRevA.87.022106,JPhysB.50.184003,JChemPhys.152.154101}

While the explicit inclusion of the light--matter coupling is typically accompanied by a perturbative treatment of the excitation--environment coupling, there are also studies concentrating on a (numerically) exact treatment of the latter, at the expense of a less transparent inclusion of the former.~\cite{JPhysChemLett.3.3136,JPhysB.51.054002,JChemPhys.151.034114} When the semiclassical description of light--matter interaction is appropriate, the time-dependent electric field can be straightforwardly incorporated in the HEOM formalism,~\cite{Kato-Tanimura-chapter-2018,JChemPhys.151.034114} whose relation to the spectroscopic picture of sequential interactions with light is not manifest. When the quantum description of light--matter interaction is in place,~\cite{JPhysChemLett.3.3136,JPhysB.51.054002} the interaction with the radiation is treated from the standpoint of quantum optics, using the so-called hybrid master-equation--HEOM approach.~\cite{JChemTheorComput.7.2166} In essence, the interaction with radiation appears in form of Markovian corrections to HEOM equations. In the end, there are also studies that propose a numerically exact treatment of both the couplings to environment and radiation,~\cite{Olsina:2014} which, however, comes with a complex formalism and huge computational costs.

In this work, we build on results of Ref.~\onlinecite{NewJPhys.12.065044}. Our approach is based on the two cornerstones of the theory of photosynthetic excitons.~\cite{Photosynthetic-excitons-book} Section~\ref{SSec:model-H} introduces the Frenkel exciton model of a molecular light-harvesting aggregate. In Sec.~\ref{Sec:heom-arb-e-t}, we shed new light on the existing approaches~\cite{Kato-Tanimura-chapter-2018,JChemPhys.151.034114} to include the interaction with pulsed laser fields into the HEOM formalism. We realize that there is a close connection between the space on which the EET dynamics has to be formulated and the maximum order up to which the interaction with the exciting field has to be included. This is very similar to the situation in the nonlinear response-function theory.~\cite{Mukamel-book} We obtain a new form of HEOM that explicitly includes the interaction with pulsed laser fields up to the second order in the field, which is fully consistent with the single-exciton Frenkel Hamiltonian commonly employed in the study of light-induced coherent EET. The analysis of Sec.~\ref{Sec:heom-arb-e-t} is actually not limited to the second-order response, and we provide a prescription for treating laser-induced nonlinearities of arbitrary order in conjunction with a numerically exact treatment of the interaction of photoinduced electronic excitations with the environment. Section~\ref{SSec:density-matrix} presents the central result of our analysis, which is valid for weak light of arbitrary properties. There, we perform a second-order treatment of the light--matter coupling and a nonperturbative treatment of the excitation--environment coupling to obtain an expression for weak light-induced excitonic dynamics that is manifestly related to the spectroscopic picture, and fully includes the dynamical interplay between nonequilibrium electronic dynamics and environmental reorganization processes. The exact result that we obtain does not allow easy analytical manipulations, and we demonstrate how it can be recast as HEOM, both in the case of semiclassical (Sec.~\ref{Sec:weak-coh}, which actually rederives the second-order results of Sec.~\ref{Sec:heom-arb-e-t}) and quantum (Sec.~\ref{Sec:weak-incoh}) treatments of the interaction with light. In addition, we analytically solve for the environmental reorganization dynamics triggered by a delta-like excitation of a single molecule (Sec.~\ref{SSec:spin-boson}), and relate our results to existing approaches, such as the Redfield theory with photoexcitation (Sec.~\ref{SSec:redfield-photo}), the nonequilibrium F\"{o}rster theory (Sec.~\ref{SSec:Foerster}), and hybrid Born--Markov--HEOM approaches (Sec.~\ref{Sec:weak-incoh}). These discussions further emphasize the advantages of our method, which are once again summarized in the concluding Sec.~\ref{Sec:conclusion}.  

\section{Model Hamiltonian}
\label{SSec:model-H}
The system of interest consists of a molecular aggregate $M$ that is in contact with the thermal bath $B$ representing its environment and with the radiation $R$. The total Hamiltonian reads as
\begin{equation}
\label{Eq:def-H-tot}
 H=H_M+H_B+H_R+H_{M-B}+H_{M-R}.
\end{equation}

The electronic excitations of the aggregate are described within the Frenkel exciton model~\cite{May-Kuhn-book,Photosynthetic-excitons-book,Valkunas-Abramavicius-Mancal-book,Agranovich-book}
\begin{equation}
\label{Eq:def-H-M-2nd-quant}
H_M=\sum_j \varepsilon_j B_j^\dagger B_j+\sum_{jk}J_{jk} B_{j}^\dagger B_{k}.
\end{equation}
In Eq.~\eqref{Eq:def-H-M-2nd-quant}, $\varepsilon_j$ are the so-called site energies, while $J_{jk}$ are resonance couplings (we take $J_{kk}=0$). The operators $B_j$ and $B_j^\dagger$ describe the destruction and creation of an excitation on site $j$, respectively, and they obey Paulion commutation relations.~\cite{May-Kuhn-book,Agranovich-book,RevModPhys.70.145}

The environment is assumed to be composed of sets of independent harmonic oscillators associated to each site
\begin{equation}
\label{Eq:H-B}
H_B=\sum_{j\xi}\hbar\omega_\xi b_{j\xi}^\dagger b_{j\xi}.
\end{equation}
The oscillators are labeled by site index $j$ and mode index $\xi$ and phonon creation and annihilation operators $b_{j\xi}^\dagger$ and $b_{j\xi}$ satisfy Bose commutation relations. The interaction of aggregate excitations and the environment is taken to be linear in mode displacements and local to each chromophore (Holstein-like coupling~\cite{AnnPhys.8.325})
\begin{equation}
H_{M-B}=\sum_{j\xi}B_j^\dagger B_j g_{j\xi}\left(b_{j\xi}^\dagger+b_{j\xi}\right)\equiv\sum_j B_j^\dagger B_j u_j,
\end{equation}
where $u_j$ is the collective environment coordinate associated with chromophore $j$. The coupling constants $g_{j\xi}$ may be related to the displacement of the equilibrium configuration of mode $\xi$ between the ground and excited electronic state of chromophore $j$.~\cite{May-Kuhn-book,Valkunas-Abramavicius-Mancal-book}

The coupling between aggregate excitations and the radiation is taken in the dipole and rotating-wave approximations
\begin{equation}
\label{Eq:H-M-R-def-quantum}
H_{M-R}=-\boldsymbol{\mu}_{eg}\cdot\mathbf{E}^{(+)}-\boldsymbol{\mu}_{ge}\cdot\mathbf{E}^{(-)}.
\end{equation}
The dipole-moment operator $\boldsymbol{\mu}$ is assumed to be a purely electronic operator $\boldsymbol{\mu}=\sum_j\mathbf{d}_j\left(B_j^\dagger+B_j\right)=\boldsymbol{\mu}_{eg}+\boldsymbol{\mu}_{ge}$, where transition dipole moment $\mathbf{d}_j$ of chromophore $j$ does not depend on environmental coordinates (Condon approximation), part $\boldsymbol{\mu}_{eg}$ contains only operators $B^\dagger$, while $\boldsymbol{\mu}_{ge}$ contains only operators $B$. $\mathbf{E}^{(\pm)}$ denotes the positive- and negative-frequency parts of the (time-independent) operator of the (transversal) electric field, so that we treat both electronic excitations and the radiation generating them on quantum level.

We assume that, at the initial instant $t_0$ of our dynamics, the total statistical operator $W(t_0)$ representing the state of the combined system of aggregate excitations, environment, and radiation, can be factorized as follows
\begin{equation}
 \label{Eq:W-tot-t0}
 W(t_0)=|g\rangle\langle g|\otimes\rho_B^g\otimes\rho_R.
\end{equation}
In Eq.~\eqref{Eq:W-tot-t0}, the aggregate is taken to be initially unexcited, the state of the environment $\rho_B^g$ is adapted to the collective electronic ground state $|g\rangle$ of the aggregate ($T=(k_B\beta)^{-1}$ is the temperature),
\begin{equation}
\label{Eq:rho-B-g}
\rho_B^g=\frac{\exp\left(-\beta H_B\right)}{\mathrm{Tr}_B\exp\left(-\beta H_B\right)},
\end{equation}
while $\rho_R$ describes the state of the radiation.

\section{Equations of Motion: Semiclassical Treatment of Light--Matter Interaction}
\label{Sec:heom-arb-e-t}
The excitation by an arbitrary time-dependent (classical) electric field $\boldsymbol{\mathcal{E}}(t)$ can be incorporated into the HEOM formalism by taking that the total purely electronic Hamiltonian is $H_M+H_{M-R}(t)$.~\cite{JChemPhys.151.034114,Kato-Tanimura-chapter-2018} Here, $H_{M-R}(t)$ is obtained from $H_{M-R}$ in Eq.~\eqref{Eq:H-M-R-def-quantum} by replacing electric-field operators $\mathbf{E}^{(\pm)}$ by the corresponding time-dependent quantities $\boldsymbol{\mathcal{E}}^{(\pm)}(t)$. Indeed, all the steps in the derivation conducted in Ref.~\onlinecite{JChemPhys.130.234111} can be repeated to obtain equations of motion for the reduced DM (RDM) $\rho(t)\equiv\sigma_\mathbf{0}(t)$ and auxiliary DMs (ADMs) $\sigma_\mathbf{n}(t)$. ADMs are fully specified by vector $\mathbf{n}$ of non-negative integers $n_{j,m}$
\begin{equation}
\label{Eq:vec-n-general-form}
 \mathbf{n}=\left\{\underbrace{(n_{0,0},n_{0,1},\dots)}_{\mathbf{n}_0},\dots,\underbrace{(n_{N-1,0},n_{N-1,1},\dots)}_{\mathbf{n}_{N-1}}\right\}.
\end{equation}
The index $j=0,\dots,N-1$ enumerates chromophores, while index $m$, in principle, does not have an upper limit and is related to the following expansion of the bath correlation function in terms of exponentially decaying factors ($t>0$)
\begin{equation}
\label{Eq:C-j-in-exp-decay}
C_j(t)=\mathrm{Tr}_B\left\{u_j^{(I)}(t) u_j(0)\rho_B^g\right\}=\sum_m c_{j,m}\:\e^{-\mu_{j,m}t}.
\end{equation}
The time dependence of the collective coordinate $u_j^{(I)}(t)$ in Eq.~\eqref{Eq:C-j-in-exp-decay} is with respect to the free-phonon Hamiltonian, Eq.~\eqref{Eq:H-B}. While expansion coefficients $c_{j,m}$ may be complex, the decay rates $\mu_{j,m}$ are assumed to be real and positive. We note that, apart from the exponential decomposition scheme [Eq.~\eqref{Eq:C-j-in-exp-decay}] adopted in this work, there are other decompositions of $C_j(t)$ from which a HEOM approach may be derived.~\cite{JChemPhys.150.244104} The bath correlation function is commonly expressed in terms of the so-called spectral density $J_j(\omega)$,
\begin{equation}
\label{Eq:C-j-in-J-j}
\begin{split}
C_j(t)&=\frac{\hbar}{\pi}\int_{-\infty}^{+\infty}\dif\omega\:J_j(\omega)\frac{\e^{\im\omega t}}{\e^{\beta\hbar\omega}-1},
\end{split}
\end{equation}
which conveniently combines information on the density of environmental-mode states and the respective coupling strengths to electronic excitations.~\cite{May-Kuhn-book,Valkunas-Abramavicius-Mancal-book}

The equation of motion for ADM $\sigma_\mathbf{n}(t)$ reads as~\cite{JChemPhys.130.234111}
\begin{equation}
\label{Eq:heom-general}
 \begin{split}
  &\partial_t\sigma_\mathbf{n}(t)=-\frac{\im}{\hbar}\left[H_M,\sigma_\mathbf{n}(t)\right]\\&+\frac{\im}{\hbar}\left[\boldsymbol{\mu}_{eg}\boldsymbol{\mathcal{E}}^{(+)}(t)+\boldsymbol{\mu}_{ge}\boldsymbol{\mathcal{E}}^{(-)}(t),\sigma_\mathbf{n}(t)\right]\\&-
  \left(\sum_j\sum_m n_{j,m}\mu_{j,m}\right)\sigma_\mathbf{n}(t)
  +\im\sum_j\sum_m\left[V_j,\sigma_{\mathbf{n}_{j,m}^+}(t)\right]\\&+\im\sum_j\sum_m n_{j,m}\left(\frac{c_{j,m}}{\hbar^2}\:V_j\sigma_{\mathbf{n}_{j,m}^-}(t)-
  \frac{c_{j,m}^*}{\hbar^2}\sigma_{\mathbf{n}_{j,m}^-}(t)V_j\right),
 \end{split}
\end{equation}
where $V_j=B_j^\dagger B_j$. Since the coupling to the radiation is explicitly included in the electronic Hamiltonian, the HEOM in Eq.~\eqref{Eq:heom-general} treats nonperturbatively not only the interaction with the bath, as usually, but also that with light. Keeping in mind that our formulation of the model Hamiltonian supports states with an arbitrary number of excitations, the result embodied in Eq.~\eqref{Eq:heom-general} is quite general. In principle, it can describe in great detail various nonlinear effects (nonlinear with respect to the exciting electric field). However, once we fix the highest order in the electric field we are interested in, there will be many elements of the DMs that do not contribute to the optical response up to that order. In other words, solving coupled equations~\eqref{Eq:heom-general} as they stand, we obtain much more information than necessary to reconstruct the optical response up to a given order. Moreover, we lack the intuitive physical picture characteristic of nonlinear spectroscopy, which is in terms of Liouville pathways, block structure of the statistical operator and evolution superoperator, etc.~\cite{Mukamel-book,Mancal-chapter-2014} In order to circumvent these deficiencies, it is enough to make a projection of the dynamics on relevant excitonic subspaces. The second-order response is fully characterized by the reduction to the subspace that can accommodate at most one excitation. This is discussed in greater detail further in this section and in Sec. SI of the Supplementary Material. Practically, the appropriate reduction to obtain the second-order response consists in the following replacements in the model Hamiltonian
\begin{equation}
\label{Eq:low-density-replacements}
B_j\to|g\rangle\langle j|,\quad B_j^\dagger\to|j\rangle\langle g|,\quad B_j^\dagger B_k\to|j\rangle\langle k|.
\end{equation}
In Eq.~\eqref{Eq:low-density-replacements}, $|j\rangle$ is the collective singly excited state featuring a selective excitation of site $j$. 

Therefore, to obtain the second-order response, we should calculate the expectation values $n_{g,\mathbf{n}}(t)\equiv\langle g|\sigma_\mathbf{n}(t)|g\rangle$, $y_{e,\mathbf{n}}(t)\equiv\langle e|\sigma_\mathbf{n}(t)|g\rangle$ and $n_{\bar e e,\mathbf{n}}(t)\equiv\langle e|\sigma_\mathbf{n}(t)|\bar e\rangle$, where $\{|e\rangle\}$ is an arbitrary basis of singly excited states (the notation is similar to that in Ref.~\onlinecite{RevModPhys.70.145}). If $\mathbf{n}=\mathbf{0}$, these three expectation values respectively represent the ground-state population, optical coherences, and singly excited-state populations and intraband coherences. Since the electronic subsystem starts from $|g\rangle\langle g|$, and since the light--matter coupling $H_{M-R}$ is the only part of the Hamiltonian that can cause transitions from the ground state to singly excited states, the following scaling relations hold~\cite{Axt-Mukamel-IMA-Chapter}
\begin{subequations}
\label{Eq:scaling-2nd}
\begin{eqnarray}
n_{g,\mathbf{n}}(t)=\delta_{\mathbf{n},\mathbf{0}}+\sum_{k=1}^{+\infty}n_{g,\mathbf{n}}^{(2k)}(t),\: n_{g,\mathbf{n}}^{(2k)}(t)\propto\mathcal{E}^{2k},\label{Eq:scaling-gg}
\\
y_{e,\mathbf{n}}(t)=\sum_{k=0}^{+\infty}y_{e,\mathbf{n}}^{(2k+1)}(t),\:y_{e,\mathbf{n}}^{(2k+1)}(t)\propto\mathcal{E}^{2k+1},\label{Eq:scaling-ge}
\\
n_{\bar e e,\mathbf{n}}(t)=\sum_{k=1}^{+\infty}n_{\bar e e,\mathbf{n}}^{(2k)}(t),\:n_{\bar e e,\mathbf{n}}^{(2k)}(t)\propto\mathcal{E}^{2k}.\label{Eq:rescaling-ee}
\end{eqnarray}
\end{subequations}
In other words, optical coherences are dominantly linear in the applied field, while excited-state populations are at least quadratic in the applied field. The environmental assistance, which actually enters through vector $\mathbf{n}$,~\cite{JChemPhys.137.194106} does not affect the scaling laws~\eqref{Eq:scaling-2nd}.~\cite{PhysRevB.53.7244}

Formulating equations of motion for $y_{e,\mathbf{n}}(t)$ and $n_{\bar e e,\mathbf{n}}(t)$ actually enables us to formulate operator equations for sectors $eg$ and $ee$ of $\sigma_\mathbf{n}(t)$. Namely, using Eqs.~\eqref{Eq:scaling-2nd} and keeping only terms that are at most of the second order in the applied field, we form the following equations for the $eg$ sector $\sigma_{eg,\mathbf{n}}(t)$ and for the $ee$ sector $\sigma_{ee,\mathbf{n}}(t)$
\begin{equation}
\label{Eq:y-x-heom-semiclassical}
 \begin{split}
  \partial_t\sigma_{eg,\mathbf{n}}(t)&=-\frac{\im}{\hbar}[H_M,\sigma_{eg,\mathbf{n}}(t)]\\
  &-\left(\sum_j\sum_m n_{j,m}\mu_{j,m}\right)\sigma_{eg,\mathbf{n}}(t)\\
  &+\delta_{\mathbf{n},\mathbf{0}}\frac{\im}{\hbar}\boldsymbol{\mathcal{E}}^{(+)}(t)\boldsymbol{\mu}_{eg}\\
  &+\im\sum_j\sum_m V_j\sigma_{eg,\mathbf{n}_{j,m}^+}(t)\\
  &+\im\sum_j\sum_m n_{j,m}\frac{c_{j,m}}{\hbar^2}\:V_j\sigma_{eg,\mathbf{n}_{j,m}^-}(t),
 \end{split}
\end{equation}
\begin{equation}
\label{Eq:n-barx-x-heom-semiclassical}
 \begin{split}
  &\partial_t\sigma_{ee,\mathbf{n}}(t)=-\frac{\im}{\hbar}\left[H_M,\sigma_{ee,\mathbf{n}}(t)\right]\\
  &-\left(\sum_j\sum_m n_{j,m}\mu_{j,m}\right)\sigma_{ee,\mathbf{n}}(t)\\
  &+\frac{\im}{\hbar}\boldsymbol{\mathcal{E}}^{(+)}(t)\boldsymbol{\mu}_{eg}\sigma_{eg,\mathbf{n}}^\dagger(t)-
  \frac{\im}{\hbar}\sigma_{eg,\mathbf{n}}(t)\boldsymbol{\mu}_{ge}\boldsymbol{\mathcal{E}}^{(-)}(t)\\
  &+\im\sum_j\sum_m\left[V_j,\sigma_{ee,\mathbf{n}_{j,m}^+}(t)\right]\\
  &+\im\sum_j\sum_m n_{j,m}\frac{c_{j,m}}{\hbar^2}\:V_j\sigma_{ee,\mathbf{n}_{j,m}^-}(t)\\&-\im\sum_j\sum_m n_{j,m}
  \frac{c_{j,m}^*}{\hbar^2}\sigma_{ee,\mathbf{n}_{j,m}^-}(t)V_j,
 \end{split}
\end{equation}
where now $V_j\to|j\rangle\langle j|$.
By reducing our dynamics to the subspace containing at most one excitation, we transform Eq.~\eqref{Eq:heom-general} into coupled equations describing evolution of optical coherences [Eq.~\eqref{Eq:y-x-heom-semiclassical}] and excited-state populations and intraband coherences [Eq.~\eqref{Eq:n-barx-x-heom-semiclassical}]. The crucial step in the transformation is the application of scaling laws in Eqs.~\eqref{Eq:scaling-2nd}, which ensure that our dynamics is consistently up to the second order in the exciting field.

Instead of the path we have taken, one could have started from the model Hamiltonian in which the low-density replacements of Eq.~\eqref{Eq:low-density-replacements} are performed, and solved Eq.~\eqref{Eq:heom-general} without ever considering the scaling laws in Eqs.~\eqref{Eq:scaling-2nd}. In that case, one would in principle obtain the solution that is exact to all orders in the exciting field. However, this exactness is only apparent, because the proper treatment of higher orders in the exciting field requires enlarging the space on which the Hamiltonian is formulated, as we discuss in more detail in Sec. SI of the Supplementary Material. Temporal evolutions of higher-order sectors of the DM (which are not taken into account) would then influence evolutions of optical coherences, excited-state populations, and intraband coherences. For example, as demonstrated in Ref.~\onlinecite{RevModPhys.70.145}, already in the third order in the electric field, equations of motion for optical coherences are coupled to equations of motion for biexcitonic amplitudes (coherences between the ground state and doubly excited states), meaning that a separate equation governing the evolution of $|jk\rangle\langle g|$ block of $\sigma_\mathbf{n}(t)$ has to be formulated. This discussion emphasizes that, once we treat the photogeneration step explicitly, we should be aware of the close connection between the largest order in the exciting field we include and the space on which the dynamics has to be formulated.
Should we limit ourselves to the Frenkel Hamiltonian for the singly excited states and, at the same time, explicitly describe the excitation generation by light, we should do that only up to the second order in the applied field.

The presented framework can be considered as a DM equivalent of the response-function approach adopted in the theory of nonlinear spectroscopy.~\cite{Mukamel-book} Our focus is on obtaining temporal evolution of various DM elements for a given waveform of the exciting electric field. This is different from the computation of nonlinear response functions, which represent the response of the system to a series of delta-like excitations. Nevertheless, there are two common assumptions underlying both our DM and response-function computations: (1) the electronic system is initially unexcited, and (2) the number of interactions with the exciting field completely determines the excitonic subspace on which the computations have to be performed.
Previous computations of coherent EET dynamics under the influence of laser fields~\cite{NewJPhys.16.053033} were practically limited to pulses of certain shapes due to the complications brought about by the time-dependent driving. On the other hand, our approach is valid for arbitrarily shaped laser pulses, and its treatment of the time-dependent driving is intuitive and consistently keeps track of interactions up to the second order.

The presented framework can be generalized to include processes that are of higher orders in the laser field. To that end, we recall the central theorem of the DCT scheme mentioned in the introduction (for more details, see Ref.~\onlinecite{RevModPhys.70.145} and references therein) guarantees that the expectation value of the normal-ordered product of $n_B$ excitonic operators $B^\dagger,B$ with respect to any $\sigma_\mathbf{n}(t)$ entering Eq.~\eqref{Eq:heom-general} is at least of the order $n_B$ in the applied laser field, i.e., $\mathrm{Tr}_M\left\{\underbrace{B_{j_1}^\dagger\dots B_{j_{n_B}}}_{n_B}\sigma_\mathbf{n}(t)\right\}=\mathcal{O}(\mathcal{E}^{n_B})$. The scaling relations [Eq.~\eqref{Eq:scaling-2nd}], which are valid for the second-order dynamics, thus represent a particular instance of the more general DCT scaling relations. This formulation of the central theorem of the DCT scheme is somewhat different from the original one in that an arbitrary-order environmental assistance of the original formulation is replaced by the expectation value with respect to an arbitrary ADM $\sigma_\mathbf{n}(t)$. The formulation presented in this paragraph relies on the results of Ref.~\onlinecite{JChemPhys.137.194106}, which provide a formal correspondence between the environmental assistance of order $2n_E$ and the HEOM's ADMs on level $n_E$. Therefore, while the DCT scheme has been typically used to study optical field-induced processes in conjunction with a perturbative treatment of the interaction with the environment,~\cite{RevModPhys.70.145,RevModPhys.74.895,RepProgPhys.67.433,PhysRevB.53.7244,PhysRevB.95.075308,JPhysChemC.121.19602,PhysRevB.65.035303} the results presented in this section open up the possibility to simultaneously study arbitrary nonlinear effects induced by arbitrarily time-varying optical fields and yet treat the interaction with the environment in a numerically exact manner. The practical procedure may be summarized as follows. For a given order $n_B$ in the exciting field, one formulates equations of motion for all possible expectation values of $n_B$ normally ordered excitonic operators starting from Eq.~\eqref{Eq:heom-general}. It may happen that some of these expectation values actually do not contribute to the optical response up to order $n_B$. For example, the central theorem of the DCT scheme predicts that the biexcitonic amplitudes $\mathrm{Tr}_M\left\{B_{j_1}B_{j_2}\sigma_\mathbf{n}(t)\right\}$ should contribute to the second-order response. However, on closer inspection, these quantities are completely decoupled from equations of motion for the expectation values of $B_j$ (optical coherences) and $B_{j_1}^\dagger B_{j_2}$ (excited-state populations and intraband coherences) and it turns out that they contribute to the third-order optical response. The details of the model Hamiltonian combined with the specific Paulion statistics of excitonic creation and annihilation operators may thus lower the number of expectation values that contribute to the optical response up to any given order. Using the ideas presented in Sec.~SI of the Supplementary Material, one can then determine the subspace on which the laser-induced excited-state dynamics has to be formulated. Further analysis is beyond the scope of this work, in which we concentrate on the weak-light second-order treatment. For the generalization of our approach to the third-order dynamics, we refer the reader to Sec.~VII of Ref.~\onlinecite{RevModPhys.70.145}.

Let us conclude this section by noting that the results we have presented so far rely heavily on the form of the light--matter interaction Hamiltonian in the semiclassical approximation. If we want to treat light quantum mechanically, too, the results of Ref.~\onlinecite{NewJPhys.12.065044} suggest that, up to the second order in the exciting field, the only information we need about light is its first-order (two-point) correlation function (indices $i,j$ label Cartesian components of a vector)
\begin{equation}
\label{Eq:G-1-ij-21-general}
G^{(1)}_{ij}(\tau_2,\tau_1)=\mathrm{Tr}_R\left\{
\left\{\mathbf{E}^{(-)}(\tau_2)\right\}_i
\left\{\mathbf{E}^{(+)}(\tau_1)\right\}_j
\rho_R
\right\}.    
\end{equation}
In the developments presented up to now, such a quantity does not directly enter Eqs.~\eqref{Eq:heom-general},~\eqref{Eq:y-x-heom-semiclassical}, and~\eqref{Eq:n-barx-x-heom-semiclassical}. However, for (classical, transform-limited) pulses, this correlation function factorizes into products of expectation values of single electric-field operators, which define classical values of the electric field~\cite{PhysRev.131.2766,PhysRevLett.114.213601,PhysRevA.91.063813}
\begin{subequations}
\label{Eq:def-G1-ij-21}
\begin{eqnarray}
G^{(1)}_{ij}(\tau_2,\tau_1)=\mathcal{E}_i^{(-)}(\tau_2)\mathcal{E}_j^{(+)}(\tau_1),\label{Eq:def-G1-ij-21-1}
\\
\mathcal{E}_i^{(\pm)}(\tau)=\mathrm{Tr}_R\left\{\left\{\mathbf{E}^{(\pm)}(\tau)\right\}_i\rho_R\right\}.\label{Eq:def-G1-ij-21-2}
\end{eqnarray}
\end{subequations}
As will be demonstrated in more detail in Sec.~\ref{Sec:weak-coh}, it is precisely this factorization that enables us to formulate Eqs.~\eqref{Eq:y-x-heom-semiclassical} and~\eqref{Eq:n-barx-x-heom-semiclassical} as they stand.

\section{General Theory of Weak Light-Induced Dynamics}
\label{SSec:density-matrix}
This section presents the central result of our exact description of the dynamics triggered by weak light of arbitrary properties. While the derivation is elementary in all its steps, it is cumbersome and thus presented in Sec. SII of the Supplementary Material. Here, we only analyze the final result for the reduced excited-state density matrix

\begin{equation}
\label{Eq:final-rho-ee-formal}
\begin{split}
\rho_{ee}^{(I)}(t)&=\int_{t_0}^t\dif\tau_2\int_{t_0}^{\tau_2}\dif\tau_1\:
 \overrightarrow{\mathcal{U}}_\mathrm{red}^{(I)}(t,\tau_2,\tau_1)A^{(I)}(\tau_2,\tau_1)+\\
 &+\int_{t_0}^t\dif\tau_2\int_{t_0}^{\tau_2}\dif\tau_1\:
 A^{(I)\dagger}(\tau_2,\tau_1)\overleftarrow{\mathcal{U}}_\mathrm{red}^{(I)}(t,\tau_2,\tau_1).
\end{split}
\end{equation}
In Eq.~\eqref{Eq:final-rho-ee-formal}, superscript $(I)$ denotes the interaction picture with respect to $H_M$, $\tau_1$ and $\tau_2$ are the instants at which the interaction with the radiation occurs, and the purely electronic operator $A^{(I)}(\tau_2,\tau_1)$ reads as
\begin{equation}
\label{Eq:def-A-I-21}
\begin{split}
 A^{(I)}(\tau_2,\tau_1)&=
 \frac{1}{\hbar^2}\sum_{i,j}G^{(1)}_{ij}(\tau_2,\tau_1)\times\\
 &\times\left\{\boldsymbol{\mu}_{eg}^{(I)}(\tau_1)\right\}_j|g\rangle\langle g|\left\{\boldsymbol{\mu}_{ge}^{(I)}(\tau_2)\right\}_i.
\end{split}
\end{equation}
The arrow above the reduced propagator sign indicates the direction of its action on the corresponding operator. The reduced propagator acting on the right reads as ($T$ is the chronological time-ordering sign)
\begin{widetext}
\begin{subequations}
\label{Eq:U-right-t-2-1-all}
\begin{align}
\overrightarrow{\mathcal{U}}_\mathrm{red}^{(I)}(t,\tau_2,\tau_1)=T\exp\left[\overrightarrow{\mathcal{W}}_c(\tau_2,\tau_1)+\overrightarrow{\mathcal{W}}_p(t,\tau_2)+\overrightarrow{\mathcal{W}}_{c-p}(t,\tau_2,\tau_1)\right],\label{Eq:U-right-t-2-1}
\\
\overrightarrow{\mathcal{W}}_c(\tau_2,\tau_1)=-\frac{1}{\hbar^2}\sum_j\int_{\tau_1}^{\tau_2}\dif s_2\int_{\tau_1}^{s_2}\dif s_1\:V_j^{(I)}(s_2)^C\:C_j(s_2-s_1)\:V_j^{(I)}(s_1)^C,\label{Eq:basic-coh}
\\
\overrightarrow{\mathcal{W}}_p(t,\tau_2)=-\frac{1}{\hbar^2}\sum_j\int_{\tau_2}^{t}\dif s_2\int_{\tau_2}^{s_2}\dif s_1\:V_j^{(I)}(s_2)^\times\left(C_j^r(s_2-s_1)\:V_j^{(I)}(s_1)^\times+\im\:C_j^i(s_2-s_1)\:V_j^{(I)}(s_1)^\circ\right),\label{Eq:basic-pops}
\\
\overrightarrow{\mathcal{W}}_{c-p}(t,\tau_2,\tau_1)=-\frac{1}{\hbar^2}\sum_j\int_{\tau_2}^t\dif s_2\int_{\tau_1}^{\tau_2}\dif s_1\:V_j^{(I)}(s_2)^\times\:C_j(s_2-s_1)\:V_j^{(I)}(s_1)^C.\label{Eq:basic-pops-cohs}
\end{align}
\end{subequations}
\end{widetext}
In Eq.~\eqref{Eq:basic-pops}, $C_j^{r/i}$ denote the real and imaginary part of the bath correlation function $C_j$ [Eq.~\eqref{Eq:C-j-in-exp-decay}], whereas the action of hyperoperators $\displaystyle{V_j^{\times,\circ,C}}$ on an operator $O$ is defined as
$V_j^\times O=[V_j,O]$,
$V_j^\circ O=\{V_j,O\}$,
$V_j^C O=V_jO$. A similar expression holds for the propagator acting on the left, as detailed in Sec.~SII of the Supplementary Material.

Equations~\eqref{Eq:final-rho-ee-formal}--\eqref{Eq:U-right-t-2-1-all} present an exact solution (with respect to the aggregate--environment coupling) of the dynamics of an excitonic system weakly driven by light of arbitrary properties. The principal novelty compared to a similar analysis conducted in Refs.~\onlinecite{NewJPhys.12.065044,ChemPhys.439.100} is that, here, we provide an exact expression for the reduced evolution superoperator that is compatible with the interaction with light, i.e., it explicitly depends on the interaction instants $\tau_1$ and $\tau_2$ with the radiation and the observation instant $t$. The two summands on the right-hand side of Eq.~\eqref{Eq:final-rho-ee-formal} are Hermitean adjoints of one another and they represent the two Liouville pathways from $|g\rangle\langle g|$ to $|e\rangle\langle e|$ which differ by the time order of the radiation interactions with the bra and ket.~\cite{Mukamel-book}

The RDM evolution can be conveniently represented in terms of diagrams showing how the state of electronic excitations changes due to interactions with radiation and due to absorptions and emissions of elementary environmental excitations.~\cite{ChemPhys.347.185,JPhysChem.97.12596} In this discussion, we assume that the instants $\tau_1$ and $\tau_2$ are fixed. We further focus on the first-order term of the reduced evolution superoperator [Eq.~\eqref{Eq:U-right-t-2-1}] and we also fix instants $s_1$ and $s_2$ [Eqs.~\eqref{Eq:basic-coh}--\eqref{Eq:basic-pops-cohs}] that describe a single environmentally assisted process.
\begin{figure}
    \centering
    \includegraphics{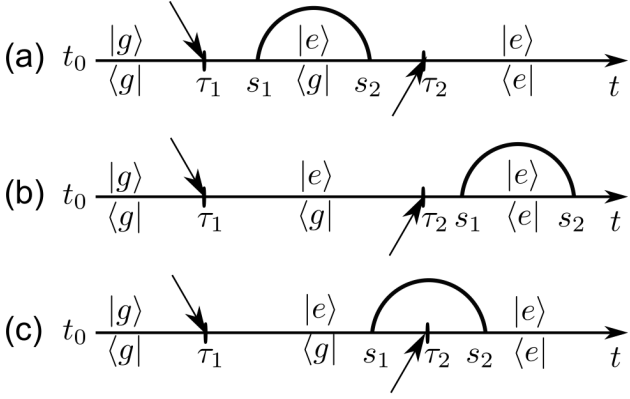}
    \caption{Primitive diagrams describing the changes that the state of the electronic system undergoes due to the interaction with the radiation and environment-assisted processes. Only the diagrams characteristic for the first-order approximation to the reduced propagator [Eq.~\eqref{Eq:U-right-t-2-1}] in the first term of Eq.~\eqref{Eq:final-rho-ee-formal} are presented. The instants $\tau_1$ and $\tau_2$ at which the electronic system interacts with light, as well as the instants $s_1$ and $s_2$ determining the environmental assistance, are fixed. The arrows at $\tau_1$ and $\tau_2$ depict interactions with light, which are reflected in changes in the ket and bra of RDM. Circumferences represent the bath correlation function $C_j(s_2-s_1)$. The observation time $t$ satisfies $t\geq\tau_2\geq\tau_1\geq t_0$. Diagram (a) corresponds to Eq.~\eqref{Eq:basic-coh}, diagram (b) corresponds to Eq.~\eqref{Eq:basic-pops}, and diagram (c) corresponds to Eq.~\eqref{Eq:basic-pops-cohs}.}
    \label{Fig:primitive-propagators}
\end{figure}
In Figs.~\ref{Fig:primitive-propagators}(a)--\ref{Fig:primitive-propagators}(c) we present the three primitive diagrams corresponding to the hyperoperators in Eqs.~\eqref{Eq:basic-coh}--\eqref{Eq:basic-pops-cohs}, respectively. The diagram in Fig.~\ref{Fig:primitive-propagators}(a) describes a single-phonon-assisted process during which the electronic subsystem is in a state of optical coherence. The diagram in Fig.~\ref{Fig:primitive-propagators}(b) describes a single-phonon-assisted process during which the electronic subsystem is entirely in the excited-state manifold. The single-phonon-assisted process represented by the primitive diagram in Fig.~\ref{Fig:primitive-propagators}(c) starts when the electronic subsystem is in a state of optical coherence, and ends when it is entirely in the excited-state manifold. There, the phonon propagator straddles two temporal sectors defined by the interactions with the radiation. These so-called straddling evolutions~\cite{ChemPhys.347.185,JPhysChem.97.12596} fully capture the nonequilibrium dynamics of the bath during different periods of photoinduced evolution.~\cite{ChemPhysLett.530.140} They are intimately connected to the quantum coherence between electronic excitations and environment and their presence is crucial to accurately describe photoinduced electronic dynamics.

Let us point out another viewpoint on the result embodied in Eq.~\eqref{Eq:final-rho-ee-formal}. Due to the assumption of the initially unexcited system, any nontrivial dynamics is ultimately induced by the interaction with the radiation because the environment alone cannot cause transitions from the ground- to the excited-state manifold. This is reflected by the fact that hyperoperator $V^C$ in Eqs.~\eqref{Eq:basic-coh} and~\eqref{Eq:basic-pops-cohs} acts after the first and before the second interaction with the radiation, while hyperoperators $V_j^{\times/\circ}$ in Eqs.~\eqref{Eq:basic-pops} and~\eqref{Eq:basic-pops-cohs}
act only after both interactions with radiation, see also Figs.~\ref{Fig:primitive-propagators}(a)--\ref{Fig:primitive-propagators}(c). Therefore, Eq.~\eqref{Eq:final-rho-ee-formal} can be reformulated by introducing a global time-ordering sign as follows
\begin{widetext}
\begin{equation}
\label{Eq:final-rho-ee-formal-hyper}
\begin{split}
&\rho^{(I)}_{ee}(t)=\int_{t_0}^t\dif\tau_2\int_{t_0}^{\tau_2}\dif\tau_1\:\frac{1}{\hbar^2}\sum_{i,j}G^{(1)}_{ij}(\tau_2,\tau_1)\times\\&\times T\left\{\exp\left[\overrightarrow{\mathcal{W}}_c(\tau_2,\tau_1)+\overrightarrow{\mathcal{W}}_p(t,\tau_2)+\overrightarrow{\mathcal{W}}_{c-p}(t,\tau_2,\tau_1)\right]\:^C\left\{\boldsymbol{\mu}_{ge}^{(I)}(\tau_2)\right\}_i\left\{\boldsymbol{\mu}_{eg}^{(I)}(\tau_1)\right\}_j^C\right\}|g\rangle\langle g|+\mathrm{H.c.}
\end{split}
\end{equation}
\end{widetext}
In Eq.~\eqref{Eq:final-rho-ee-formal-hyper}, we introduced hyperoperator $^CV$ as $^CVO=OV$, for any operators $V$ and $O$. This viewpoint will be useful in our discussion in Sec.~\ref{SSec:Foerster}, where we emphasize the similarities between the descriptions of the second-order photoexcitation process starting from the ground state and the F\"{o}rster energy transfer from an excited donor to an unexcited acceptor.

Even though the result embodied in Eqs.~\eqref{Eq:final-rho-ee-formal}--\eqref{Eq:U-right-t-2-1-all} is remarkable, it is not very useful for actual computations, principally due to the time-ordering sign that renders analytical manipulations difficult. Nevertheless, whenever the bath correlation function $C_j(t)$ can be represented in the form given in Eq.~\eqref{Eq:C-j-in-exp-decay}, Eq.~\eqref{Eq:final-rho-ee-formal} can be recast as an infinite hierarchy of equations of motion for the RDM and ADMs.~\cite{JChemPhys.130.234111} However, the details of this procedure now depend on the form of operator $A^{(I)}(\tau_2,\tau_1)$ [Eq.~\eqref{Eq:def-A-I-21}], i.e., on the temporal and statistical properties of the radiation.

\section{Excitation by Weak (Coherent) Laser Pulses}
\label{Sec:weak-coh}
As has been recently discussed in Refs.~\onlinecite{PhysRevLett.114.213601,PhysRevA.91.063813}, a pulse of light may be understood as a classical-like state of the electromagnetic field, whose energy density is localized and which can be specified by the spatial position around which it is localized, propagation direction, polarization, and spectral distribution. In essence, the quantum state representing the classical pulse whose bandwidth is determined by its spectral distribution is a coherent state which, as first realized by Glauber,~\cite{PhysRev.131.2766} factorizes the $2n$-point radiation correlation function into product of $2n$ expectation values of the electric-field operator. In particular, $G^{(1)}_{ij}(\tau_2,\tau_1)$ is then factorized as predicted by Eq.~\eqref{Eq:def-G1-ij-21}, so that $A^{(I)}(\tau_2,\tau_1)$ [Eq.~\eqref{Eq:def-A-I-21}] assumes the form
\begin{equation}
\label{Eq:A-21-semiclassical}
\begin{split}
A^{(I)}(\tau_2,\tau_1)&=
\frac{1}{\hbar^2}\left[\boldsymbol{\mu}_{eg}^{(I)}(\tau_1)\cdot\boldsymbol{\mathcal{E}}^{(+)}(\tau_1)\right]|g\rangle\langle g|\times\\&\times\left[\boldsymbol{\mu}_{ge}^{(I)}(\tau_2)\cdot\boldsymbol{\mathcal{E}}^{(-)}(\tau_2)\right].
\end{split}
\end{equation}
In other words, the result is the same as if we used the semiclassical form of the light--matter coupling from the very beginning, without any reference to electric-field operators. Therefore, further developments towards the HEOM have to result in Eqs.~\eqref{Eq:y-x-heom-semiclassical} and~\eqref{Eq:n-barx-x-heom-semiclassical} that govern time evolution of optical coherences and excited-state populations and intraband coherences, respectively.

While the hierarchy counterpart of Eq.~\eqref{Eq:final-rho-ee-formal} for the RDM in the excited-state sector is Eq.~\eqref{Eq:n-barx-x-heom-semiclassical}, our previous discussion has not dealt with the RDM counterpart of the hierarchy for optical coherences [Eq.~\eqref{Eq:y-x-heom-semiclassical}]. In Sec.~SII of the Supplementary Material, we demonstrate that the exact solution (with respect to the excitation--environment coupling) in the $eg$ sector reads as
\begin{equation}
\label{Eq:def-opt-coh}
\rho_{eg}^{(I)}(t)=\int_{t_0}^t\dif\tau\:U_\mathrm{red}^{(I)}(t,\tau)\frac{\im}{\hbar}\boldsymbol{\mu}_{eg}^{(I)}(\tau)\boldsymbol{\mathcal{E}}^{(+)}(\tau)|g\rangle\langle g|,
\end{equation}
where the reduced propagator for optical coherences reads as [see Eqs.~\eqref{Eq:U-right-t-2-1} and~\eqref{Eq:basic-coh}]
\begin{equation}
\label{Eq:red-prop-opt-cohs}
U_\mathrm{red}^{(I)}(t,\tau)=\overrightarrow{\mathcal{U}}^{(I)}_\mathrm{red}(t,t,\tau)=T\exp\left[\overrightarrow{\mathcal{W}}_c(t,\tau)\right].
\end{equation}

\begin{widetext}
The manipulations that are necessary to recast Eqs.~\eqref{Eq:def-opt-coh} and~\eqref{Eq:final-rho-ee-formal} as the HEOM presented in Eqs.~\eqref{Eq:y-x-heom-semiclassical} and~\eqref{Eq:n-barx-x-heom-semiclassical}, respectively, proceed as usually.~\cite{JChemPhys.130.234111} For the sake of completeness, here, we only present the definitions of ADMs (in the interaction picture) for optical coherences
\begin{equation}
\label{Eq:define-1st-tier-opt-coh}
\sigma_{eg,\mathbf{n}}^{(I)}(t)=\int_{t_0}^t\dif\tau\:
T\left\{\prod_j\prod_m\left[\int_{\tau}^t\dif s\:\e^{-\mu_{j,m}(t-s)}\:\im\frac{c_{j,m}}{\hbar^2}\:V_j^{(I)}(s)^C\right]^{n_{j,m}}U^{(I)}_\mathrm{red}(t,\tau)
\right\}\frac{\im}{\hbar}\boldsymbol{\mu}_{eg}^{(I)}(\tau)\boldsymbol{\mathcal{E}}^{(+)}(\tau)|g\rangle\langle g|,
\end{equation}
and for excited-state populations and intraband coherences ($c_{j,m}^{r/i}$ denote the real/imaginary part of complex coefficients $c_{j,m}$)
\begin{equation}
\label{Eq:define-1st-tier-pops-intra-cohs}
\begin{split}
\sigma_{ee,\mathbf{n}}^{(I)}(t)=\int_{t_0}^t\dif\tau_2\int_{t_0}^{\tau_2}\dif\tau_1\:T\left\{\prod_j\prod_m\left[\int_{\tau_2}^t\dif s\:\e^{-\mu_{j,m}(t-s)}\left(\im\frac{c_{j,m}^r}{\hbar^2}\:V_j^{(I)}(s)^\times-\frac{c_{j,m}^i}{\hbar^2}\:V_j^{(I)}(s)^\circ\right)+\right. \right. \\ \left. \left. +\int_{\tau_1}^{\tau_2}\dif s\:\e^{-\mu_{j,m}(t-s)}\:\im\frac{c_{j,m}}{\hbar^2}\:V_j^{(I)}(s)^C\right]^{n_{j,m}}\overrightarrow{\mathcal{U}}_\mathrm{red}^{(I)}(t,\tau_2,\tau_1)\right\}A^{(I)}(\tau_2,\tau_1)+\mathrm{H.c.}
\end{split}
\end{equation}
\end{widetext}

\begin{figure}[htbp]
    \centering
    \includegraphics{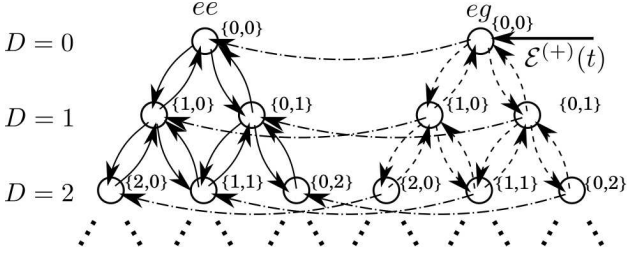}
    \caption{Schematic representation of the HEOM for optical coherences and excited-state populations and intraband coherences in the case of excitation by a weak laser pulse. For the sake of simplicity, the aggregate comprises $N=2$ chromophores and only $K=1$ term in the exponential decomposition of the bath correlation function $C_j(t)$ is taken into account. Individual DMs are represented by circles, while the driving by the electric field $\mathcal{E}^{(+)}(t)$, which directly affects only the optical-coherence RDM, is presented by the straight horizontal arrow. $D$ denotes the level of the hierarchy, and each DM is accompanied by the corresponding vector $\mathbf{n}$, see Eq.~\eqref{Eq:vec-n-general-form}. Curved dashed arrows represent hierarchical links between optical-coherence DMs, while curved solid arrows represent hierarchical links between excited-state DMs. The fact that DM $\sigma_{eg,\mathbf{n}}(t)$ acts as the source term in the EOM for $\sigma_{ee,\mathbf{n}}(t)$ is reflected in the diagram by the presence of curved dash-dotted arrows pointing from $\sigma_{eg,\mathbf{n}}(t)$ towards $\sigma_{ee,\mathbf{n}}(t)$.}
    \label{Fig:hierarchy}
\end{figure}

Before discussing the relation of the HEOM embodied in Eqs.~\eqref{Eq:y-x-heom-semiclassical} and~\eqref{Eq:n-barx-x-heom-semiclassical} to existing theories of the dynamics of electronic excitations induced by weak laser pulses, let us briefly comment on the way in which the photoexcitation enters the HEOM. The electric field explicitly enters the hierarchy for optical coherences only on the level of RDM, see Eq.~\eqref{Eq:y-x-heom-semiclassical}. Environmentally assisted optical coherences then act as source terms for environmentally assisted excited-state populations and intraband coherences, see Eq.~\eqref{Eq:n-barx-x-heom-semiclassical}. Moreover, the source term for the $ee$ sector of ADM characterized by vector $\mathbf{n}$ comprises only the $eg$ sector of ADM characterized by the same vector $\mathbf{n}$. The hierarchy is schematically presented in Fig.~\ref{Fig:hierarchy} for $N=2$ chromophores and $K=1$ terms in the decomposition of the bath correlation function $C_j(t)$ in Eq.~\eqref{Eq:C-j-in-exp-decay}.

\subsection{Impulsive Photoexcitation of Pure-Dephasing Spin--Boson Model: Analytical Results}
\label{SSec:spin-boson}
Let us now concentrate on the case of only one chromophore. The Hamiltonian, Eq.~\eqref{Eq:def-H-tot}, then reads as
\begin{equation}
\label{Eq:spin-boson}
\begin{split}
H&=\varepsilon_e|e\rangle\langle e|+\sum_\xi\hbar\omega_\xi b_\xi^\dagger b_\xi\\&+\sum_\xi g_\xi|e\rangle\langle e|\left(b_\xi^\dagger+b_\xi\right)\\
&-\mathbf{d}_{eg}\cdot\left(\boldsymbol{\mathcal{E}}^{(+)}(t)|e\rangle\langle g|+\boldsymbol{\mathcal{E}}^{(-)}(t)|g\rangle\langle e|\right).
\end{split}
\end{equation}
Equation~\eqref{Eq:spin-boson} is actually the pure-dephasing spin--boson Hamiltonian (or the independent-boson Hamiltonian, see Ref.~\onlinecite{Mahan-book}), in which $\varepsilon_e$ is the energy splitting between the two local energy levels (the ground state $|g\rangle$ and the singly excited state $|e\rangle$), and there is no tunneling between the two levels. The hyperoperators appearing in the reduced evolution superoperator [Eqs.~\eqref{Eq:U-right-t-2-1-all}] are time-independent, meaning that the time-ordering signs are not effective. This circumstance enables us to obtain analytical insights into the photoexcitation dynamics of the pure-dephasing spin--boson model in the impulsive limit.

The waveform of the positive-frequency part of the electric field is taken to be
\begin{equation}
\label{Eq:waveform-impulsive}
\boldsymbol{\mathcal{E}}^{(+)}(t)=\mathbf{e}\:\mathcal{E}_0\delta(t)\e^{-\im\Omega_p t},
\end{equation}
where vector $\mathbf{e}$ defines the polarization of the pulse, $\Omega_p$ is its central frequency, and $\mathcal{E}_0$ is its amplitude. If the initial instant is $t_0<0$, the excited-state RDM for $t>0$ reads as~\cite{May-Kuhn-book}
\begin{equation}
\label{Eq:rho-ee-spin-boson}
\rho_{ee}(t)=\frac{1}{\hbar^2}\left|\left(
\mathbf{d}_{eg}\cdot\mathbf{e}
\right)\mathcal{E}_0\right|^2|e\rangle\langle e|\equiv P_e|e\rangle\langle e|.
\end{equation}
At the same time, the optical-coherence RDM
\begin{equation}
\label{Eq:rho-eg-spin-boson}
\rho_{eg}(t)=\frac{\im}{\hbar}\left(
\mathbf{d}_{eg}\cdot\mathbf{e}
\right)\mathcal{E}_0\:\e^{-\im\varepsilon_et/\hbar}\:\e^{-g(t)}|e\rangle\langle g|
\end{equation}
exponentially decays to zero~\cite{Mancal-chapter-2014} on a time scale determined by the temporal behavior of the lineshape function
\begin{equation}
\label{Eq:def-lineshape}
g(t)=\frac{1}{\hbar^2}\int_0^t\dif s_2\int_0^{s_2}\dif s_1\:C(s_1).
\end{equation}
In a certain sense, Eqs.~\eqref{Eq:rho-ee-spin-boson} and~\eqref{Eq:rho-eg-spin-boson} formally demonstrate that the propagation scheme adopted in, e.g., Ref.~\onlinecite{JChemPhys.130.234111}, is physically sensible. Namely, optical coherences generated upon impulsive photoexcitation quickly decay to zero and, more importantly, they do not act as sources for excited-state populations and intraband coherences for $t>0$, see Eqs.~\eqref{Eq:waveform-impulsive},~\eqref{Eq:n-barx-x-heom-semiclassical} and~\eqref{Eq:y-x-heom-semiclassical}. Therefore, upon a delta-like photoexcitation, it is justified to propagate only the excited-state dynamics. The reduced propagator for the excited-state sector, Eq.~\eqref{Eq:U-right-t-2-1}, then becomes the reduced propagator used in Ref.~\onlinecite{JChemPhys.130.234111}.

\begin{figure}[htbp!]
    \centering
    \includegraphics[scale=1.1]{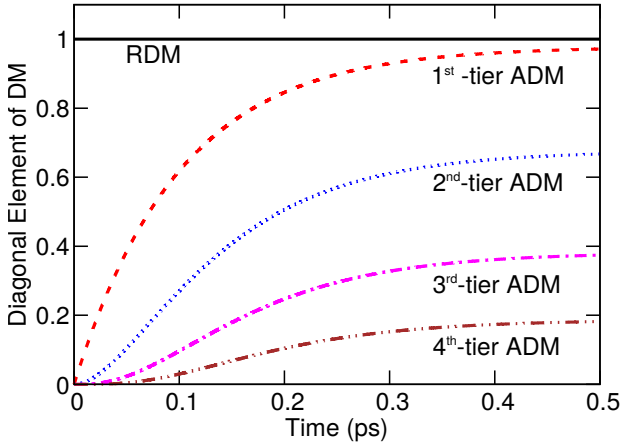}
    \caption{(Color online) Time evolution of the RDM and first four nontrivial ADMs following the impulsive excitation of the pure-dephasing spin--boson model. The spectral density of the excitation--environment interaction is assumed to be of the Drude--Lorentz type, see Eq.~\eqref{Eq:J-drude-lorentz}. The results are obtained using Eq.~\eqref{Eq:adm-spin-boson-after-rescaling} with the following values of model parameters: reorganization energy $\lambda=100\:\mathrm{cm}^{-1}$, bath relaxation time $\gamma^{-1}=100\:\mathrm{fs}$, temperature $T=300\:\mathrm{K}$.}
    \label{Fig:exact}
\end{figure}

Although the excited-state RDM does not evolve in time, the impulsive photoexcitation triggers environmental reorganization processes, whose dynamics is encoded in ADMs. Using the definition of the first-tier excited-state ADM in Eq.~\eqref{Eq:define-1st-tier-pops-intra-cohs} and specializing to the single-chromophore case and impulsive excitation, we obtain
\begin{equation}
\sigma_{ee,\mathbf{0}_m^+}(t)=-2\frac{c_m^i}{\hbar^2}\frac{1-\e^{-\mu_m t}}{\mu_m}P_e|e\rangle\langle e|.
\end{equation}
In essence, the only nontrivial contribution comes from the anticommutator with $|e\rangle\langle e|$, which produces a factor of 2. A similar analysis can be conducted for $d$th-tier ($d\geq 1$) excited-state ADM with the final result
\begin{equation}
\label{Eq:adm-ee-spin-boson-general}
\sigma_{ee,\mathbf{0}_{m_1\dots m_d}^+}(t)=
(-2)^d\prod_{p=1}^d\left(\frac{c_{m_p}^i}{\hbar^2}\:\frac{1-\e^{-\mu_{m_p}t}}{\mu_{m_p}}\right)P_e|e\rangle\langle e|.
\end{equation}
Therefore, within the pure-dephasing spin--boson model, we can analytically compute the nonequilibrium environmental dynamics initiated by a delta-like photoexcitation. The result embodied in Eq.~\eqref{Eq:adm-ee-spin-boson-general} becomes particularly interesting in the archetypal case of overdamped Brownian oscillator spectral density
\begin{equation}
\label{Eq:J-drude-lorentz}
J(\omega)=2\lambda\frac{\omega\gamma}{\omega^2+\gamma^2},
\end{equation}
when only the coefficient $c_0$ connected to the Drude pole $\mu_0=\gamma$ has an imaginary part [see also Eq.~\eqref{Eq:C-j-in-J-j}]
\begin{equation}
c_0=\lambda\cdot\hbar\gamma\left[\cot\left(\frac{\beta\hbar\gamma}{2}\right)-\im\right].
\end{equation}
In this case, the only excited-state ADMs which exhibit a nontrivial temporal evolution are the ones featuring an exclusive excitation of the Drude pole. After performing suitable rescalings, which ensure that ADMs are dimensionless and indeed decay to zero in high enough hierarchical orders,~\cite{JChemPhys.130.084105} we finally obtain for $d\geq 0$
\begin{equation}
\label{Eq:adm-spin-boson-after-rescaling}
\begin{split}
&\langle e|\sigma_{ee,\mathbf{0}_{m_1\dots m_d}^+}^\mathrm{resc}(t)|e\rangle/P_e=\delta_{m_1,0}\dots\delta_{m_d,0}\:\times\\
&\times\frac{2^d}{\sqrt{d!}}\left(\frac{\lambda}{\hbar\gamma}\right)^{d/2}\left[1+\cot^2\left(\frac{\beta\hbar\gamma}{2}\right)\right]^{-d/4}\left(1-\e^{-\gamma t}\right)^d.
\end{split}
\end{equation}
In Fig.~\ref{Fig:exact}, we present the time evolution of the RDM and first four nontrivial ADMs that is predicted by Eq.~\eqref{Eq:adm-spin-boson-after-rescaling}.
The numerical computations of the dynamics of impulsively photoexcited spin--boson model performed in Ref.~\onlinecite{JChemPhys.130.234111} (see Fig. 1 and the corresponding discussion) employed the high-temperature approximation, in which the expansion of the bath correlation function [Eq.~\eqref{Eq:C-j-in-exp-decay}] contains only the Drude contribution (term with $m=0$). Interestingly, our analytical result [Eq.~\eqref{Eq:adm-spin-boson-after-rescaling}] demonstrates that, in that case, the high-temperature approximation actually gives an exact solution.

Let us also note that the procedure outlined can be repeated to obtain optical-coherence ADMs. However, judging by Eq.~\eqref{Eq:define-1st-tier-opt-coh}, there will be no restrictions on $m$s that can be excited. This is not at variance with constraints present in Eq.~\eqref{Eq:adm-spin-boson-after-rescaling} because, in the impulsive limit, optical coherences are not sources for purely excited-state dynamics.

\subsection{Redfield Theory with Photoexcitation}
\label{SSec:redfield-photo}
Here, we discuss how, in the limit of weak excitation--environment interaction, our results for $\rho^{(I)}_{ee}(t)$ and $\rho^{(I)}_{eg}(t)$ reduce to the results of Ref.~\onlinecite{RevModPhys.70.145,PhysRevB.65.035303}, where the photoexcitation is treated up to the second order in the optical field, while the environment-induced relaxation processes are described within the Redfield theory. Our strategy is similar to the one used in Ref.~\onlinecite{JChemPhys.130.234111} to accomplish a similar goal.

If we assume that the characteristic decay time of the bath correlation function $C_j(t)$ is short compared to the time scales of the dynamics we are interested in, we can employ the Markov approximation to reduce Eqs.~\eqref{Eq:final-rho-ee-formal} and~\eqref{Eq:def-opt-coh} to a system of coupled second-order equations for the excited-state and optical-coherence sectors of the RDM.~\cite{May-Kuhn-book,Valkunas-Abramavicius-Mancal-book} The final result is commonly written in the exciton basis $\{|x\rangle\}$, defined by $H_M|x\rangle=\hbar\omega_x|x\rangle$, and assumes the form of Redfield equations with photoexcitation. The optical coherence $y_x(t)=\langle x|\rho_{eg}(t)|g\rangle$ evolves according to
\begin{equation}
\label{Eq:dif-t-y-x-red}
\begin{split}
\partial_t y_x(t)&=-\im\omega_x y_x(t)+\frac{\im}{\hbar}\boldsymbol{\mu}_x\cdot\boldsymbol{\mathcal{E}}^{(+)}(t)\\
&-\sum_{x'}\left(\sum_{\tilde x}\Gamma_{x\tilde x\tilde x x'}\right)y_{x'}(t),
\end{split}
\end{equation}
while exciton populations and interexciton coherences $n_{\bar x x}(t)=\langle x|\rho_{ee}(t)|\bar x\rangle$ obey
\begin{equation}
\label{Eq:dif-t-n-barx-x-red}
\begin{split}
\partial_t n_{\bar x x}(t)&=-\im\left(\omega_x-\omega_{\bar x}\right)n_{\bar x x}(t)-\\
&-\frac{\im}{\hbar}\boldsymbol{\mu}_{\bar x}^*\cdot\boldsymbol{\mathcal{E}}^{(-)}(t)\:y_{x}(t)+\\
&+\frac{\im}{\hbar}y_{\bar x}^*(t)\boldsymbol{\mu}_x\cdot\boldsymbol{\mathcal{E}}^{(+)}(t)-\\
&-\sum_{\bar x'x'}\mathcal{R}_{\bar x x\bar x'x'}n_{\bar x'x'}(t).
\end{split}
\end{equation}
In Eq.~\eqref{Eq:dif-t-y-x-red}, the damping matrix $\Gamma_{xx'\bar x\bar x'}$ is defined as
\begin{equation}
\begin{split}
\Gamma_{xx'\bar x\bar x'}&=
\sum_j\langle x|j\rangle\langle j|x'\rangle\langle \bar x|j\rangle\langle j|\bar x'\rangle\times\\
&\times\int_0^{+\infty}\dif s\:\frac{C_j(s)}{\hbar^2}\e^{\im(\omega_{\bar x'}-\omega_{\bar x})s},
\end{split}
\end{equation}
while the Redfield tensor $\mathcal{R}_{\bar x x\bar x'x'}$ appearing in Eq.~\eqref{Eq:dif-t-n-barx-x-red} assumes the standard form
\begin{equation}
\begin{split}
\mathcal{R}_{\bar x x\bar x'x'}&=-\Gamma_{\bar x'\bar x x x'}-\Gamma_{x' x\bar x\bar x'}^*+\\
&+\delta_{\bar x'\bar x}\sum_{\tilde x}\Gamma_{x\tilde x\tilde x x'}
+\delta_{x'x}\sum_{\tilde x}\Gamma_{\bar x\tilde x\tilde x\bar x'}^*.
\end{split}
\end{equation}
We have also introduced elements of the dipole-moment operator in the excitonic basis $\boldsymbol{\mu}_x=\langle x|\boldsymbol{\mu}_{eg}|g\rangle$. Although quite standard, the derivation of Eqs.~\eqref{Eq:dif-t-y-x-red} and~\eqref{Eq:dif-t-n-barx-x-red} from Eqs.~\eqref{Eq:final-rho-ee-formal} and~\eqref{Eq:def-opt-coh} deserves attention, and we present it in Sec.~SIII of the Supplementary Material.

In applications, it is common to neglect the imaginary parts of the Redfield tensor,~\cite{RevModPhys.70.145,May-Kuhn-book} which give rise to renormalizations of transition frequencies. However, as discussed in Ref.~\onlinecite{JChemPhys.130.234110}, this is not correct, especially when we discuss the Redfield equation without the secular approximation. Moreover, as the following discussion demonstrates, the application of Eqs.~\eqref{Eq:dif-t-y-x-red} and~\eqref{Eq:dif-t-n-barx-x-red} to describe laser-induced dynamics of electronic excitations that are strongly coupled to relatively slow nuclear motions runs into more serious difficulties than those caused by neglecting renormalizations of transition frequencies or applying the secular approximation.

\begin{figure}
 \includegraphics[scale=0.9]{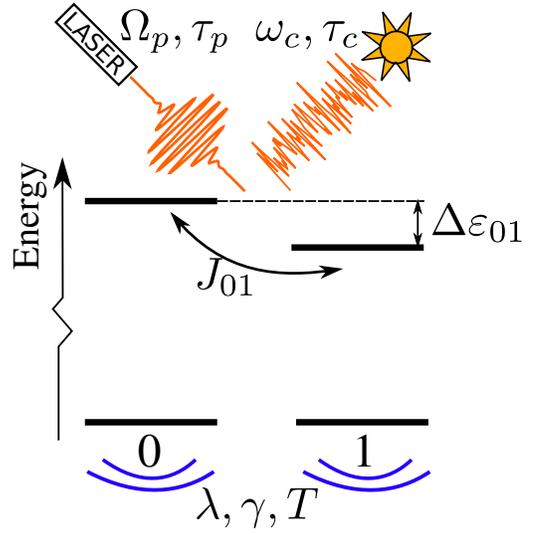}
 \caption{(Color online) Scheme of the model dimer. The difference between local energy levels is $\Delta\varepsilon_{01}=\varepsilon_0-\varepsilon_1=100\:\mathrm{cm}^{-1}$, and the electronic coupling is $J_{01}=100\:\mathrm{cm}^{-1}$. The transition dipole moment of site 1 is assumed to be perpendicular to the polarization vector of the exciting field, whereas the magnitude of the projection of the transition dipole moment of site 0 onto the polarization vector is $d_{eg}$. Each chromophore is in contact with its thermal bath (schematically represented by the motion lines below chromophore numbers) and the spectral density of the excitation--environment interaction is assumed to be the Drude--Lorentz spectral density, see Eq.~\eqref{Eq:J-drude-lorentz}, whose parameters $\gamma$ and $\lambda$ are identical on both sites. The bath relaxation time is $\gamma^{-1}=100\:\mathrm{fs}$, while the temperature is $T=300\:\mathrm{K}$. The initially unexcited dimer is excited by a weak laser pulse (characterized by the pulse central frequency $\Omega_p$ and duration $\tau_p$, see Fig.~\ref{Fig:heom-redfield}) or by weak incoherent light (characterized by the central frequency $\omega_c$ and correlation time $\tau_c$, see Figs.~\ref{Fig:full-vs-wnm} and~\ref{Fig:interexc-thermal}).}
 \label{Fig:dimer}
\end{figure}

\begin{widetext}
%\onecolumngrid

\begin{figure}[htbp!]
    \centering
    \includegraphics[scale=0.75]{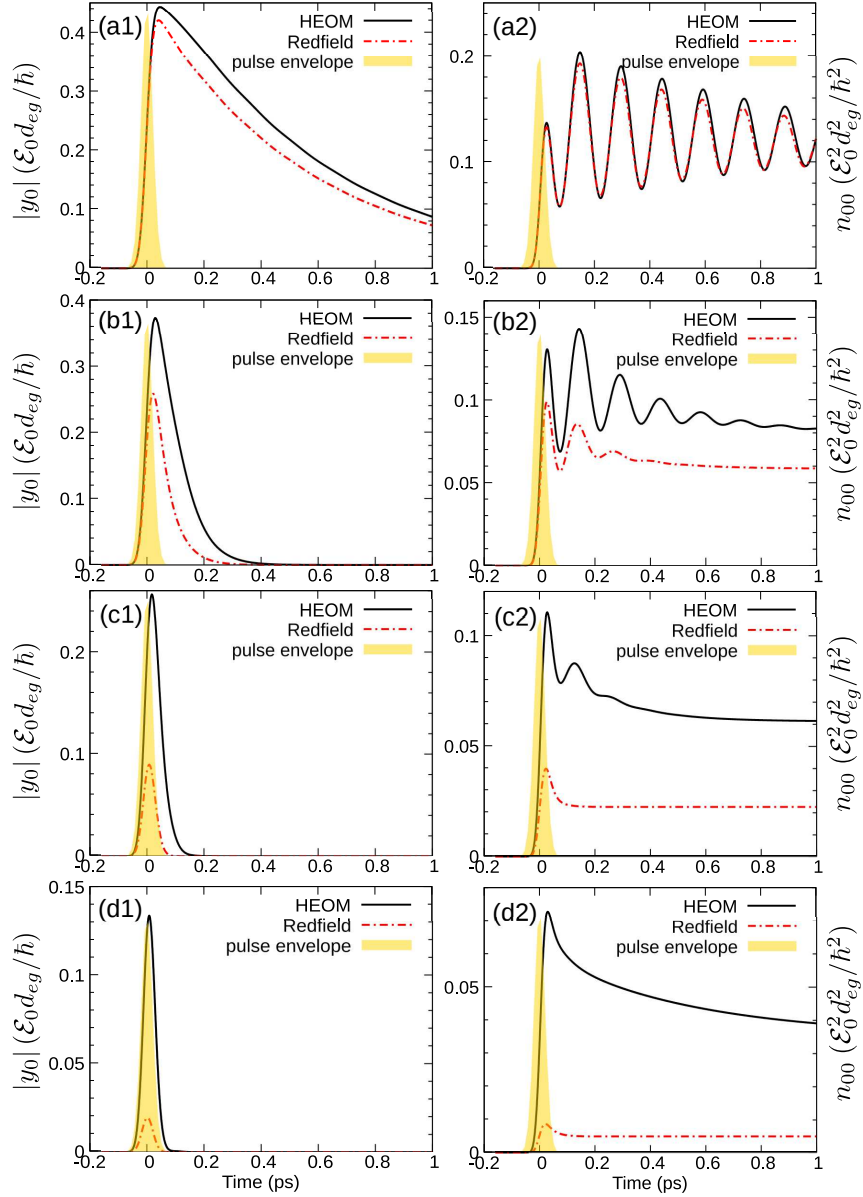}
    \caption{(Color online) Time evolution of the optical coherence modulus [(a1)--(d1)] and population [(b2)--(d2)] of site 0 following a pulsed photoexcitation of the model dimer (see Fig.~\ref{Fig:dimer}). The computation is performed using the HEOM formalism incorporating the photoexcitation [Eqs.~\eqref{Eq:y-x-heom-semiclassical} and~\eqref{Eq:n-barx-x-heom-semiclassical}, solid curves] and the Redfield theory incorporating the photoexcitation [Eqs.~\eqref{Eq:dif-t-y-x-red} and~\eqref{Eq:dif-t-n-barx-x-red}, dash-dotted curves], while the envelope of the photoexcitation is represented by shaded areas. The waveform of the excitation is $\mathcal{E}^{(+)}(t)=\mathcal{E}_0\exp\left(-\im\Omega_p t-t^2/(2\tau_p^2)\right)/(\tau_p\sqrt{2\pi})$, where the duration of the pulse is $\tau_p=20\:\mathrm{fs}$, while the central frequency $\Omega_p$ is tuned to the vertical transition frequency of site 0. The reorganization energy assumes the following values: $\lambda=2\:\mathrm{cm}^{-1}$ in (a1) and (a2), $\lambda=20\:\mathrm{cm}^{-1}$ in (b1) and (b2), $\lambda=100\:\mathrm{cm}^{-1}$ is (c1) and (c2), and $\lambda=500\:\mathrm{cm}^{-1}$ in (d1) and (d2).}
    \label{Fig:heom-redfield}
\end{figure}
%\twocolumngrid
\end{widetext}

In Figs.~\ref{Fig:heom-redfield}(a1)--\ref{Fig:heom-redfield}(d2) we compare the photoinduced electronic dynamics of a dimer (see Fig.~\ref{Fig:dimer}) treated by our HEOM formalism incorporating the photoexcitation [Eqs.~\eqref{Eq:y-x-heom-semiclassical} and~\eqref{Eq:n-barx-x-heom-semiclassical}] and the Redfield formalism incorporating the photoexcitation [Eqs.~\eqref{Eq:dif-t-y-x-red} and~\eqref{Eq:dif-t-n-barx-x-red}]. Relevant parameters of the model dimer are summarized in the caption of Fig.~\ref{Fig:dimer}.

For the weakest excitation--environment coupling, see Figs.~\ref{Fig:heom-redfield}(a1) and~\ref{Fig:heom-redfield}(a2), the results predicted by the two approaches are quite similar, as expected. However, as the excitation--environment coupling is increased, the dynamics predicted by the Redfield theory deviates both qualitatively (e.g., absence of oscillatory features) and quantitatively from the numerically exact results, see Figs.~\ref{Fig:heom-redfield}(b1)--\ref{Fig:heom-redfield}(d2). The reasons for such deviations are summarized in the following. 

Firstly, the relaxation tensor employed in Eqs.~\eqref{Eq:dif-t-y-x-red} and~\eqref{Eq:dif-t-n-barx-x-red} is time-independent, i.e., it cannot accurately capture the very first steps of the nuclear reorganization dynamics initiated by photoexcitation. In the derivation of Eqs.~\eqref{Eq:dif-t-y-x-red} and~\eqref{Eq:dif-t-n-barx-x-red}, we obtained time-local equations because we ceased to keep track of the exact instants of the interaction with light by formally setting the difference between the observation instant $t$ and the last instant of the interaction with light $\tau$ to infinity. Such an approximation is reasonable whenever the bath correlation time and/or the excitation--environment coupling are small enough. These conditions are typically satisfied in ultrafast semiconductor optics,~\cite{RevModPhys.74.895,RepProgPhys.67.433} which explains the success of methods relying on equations such as Eqs.~\eqref{Eq:dif-t-y-x-red} and~\eqref{Eq:dif-t-n-barx-x-red} to describe ultrafast semiconductor dynamics. On the other hand, in view of the intermediate regime to which photosynthetic EET belongs,~\cite{RevModPhys.90.035003,AnnuRevPhysChem.66.69,AnnuRevCondensMatterPhys.3.333} transient features of light-triggered nuclear reorganization dynamics become crucial to properly characterize electronic dynamics in photosynthetic aggregates. In other words, one has to keep track of the exact instants $\tau_1$ and $\tau_2$ of the interaction with light, which our formalism manifestly does. One may hope to partially cure the deficiencies of the dynamics predicted by Eqs.~\eqref{Eq:dif-t-y-x-red} and~\eqref{Eq:dif-t-n-barx-x-red} by replacing the time-independent Redfield tensor by its time-dependent counterpart, see, e.g., Ref.~\onlinecite{ChemPhys.404.103}. However, as argued in the Supporting Information to Ref.~\onlinecite{SciRep.6.26230}, such a replacement in a time-local equation for RDM would have to rely on the rather arbitrary instant $t_0$ in which we prescribe the initial condition [Eq.~\eqref{Eq:W-tot-t0}], which would give a reasonable description only in the limit of impulsive excitation at $t_0$. For pulses of finite duration, the correct description of ultrafast dynamics has to be on the time-nonlocal level.

Secondly, the derivation presented in the Supplementary Material suggests that Eqs.~\eqref{Eq:dif-t-y-x-red} and~\eqref{Eq:dif-t-n-barx-x-red} neglect the nonequilibrium dynamics of the bath in the period between the two interactions with light. Again, our formalism manifestly includes such dynamics through the HEOM for optical coherences. On the other hand, the change in the bath state in the period between the two interactions with the light can be partially taken into account, even on the time-local level, through the so-called photoinduced correlation term that was identified in Ref.~\onlinecite{SciRep.6.26230} (and also, in a more specialized setting, in Ref.~\onlinecite{ChemPhys.404.103}). In the language of the standard density matrix theory, the photoinduced correlation term arises from the combined action of the environmental assistance and the interaction with the exciting field. While the neglect of such a term can be justified in semiconductor optics,~\cite{RevModPhys.74.895,RepProgPhys.67.433} its effect on the dynamics may be nontrivial in the case of slow bath and/or strong excitation--environment coupling.

\subsection{Nonequilibrium Generalization of F\"{o}rster Theory}
\label{SSec:Foerster}
As pointed out in Ref.~\onlinecite{JChemPhys.130.234111}, the F\"{o}rster limit~\cite{Foerster1964} cannot be directly obtained from the analytical results presented there, simply because the initial environmental density matrix $\rho_B^g$ is assumed to describe the equilibrium of environmental modes when there are no electronic excitations in the system. However, in the following, we demonstrate how, under appropriate approximations, the results of Ref.~\onlinecite{JChemPhys.130.234111}, i.e., our results in the limit of ultrashort excitation, lead to the nonequilibrium generalization of the F\"{o}rster theory proposed in Refs.~\onlinecite{ChemPhys.275.319,JChemPhys.146.174109}.

Let us limit our discussion to an aggregate containing two chromophores, see Fig.~\ref{Fig:dimer}, one acting as the excitation donor ($D$, chromophore 0 in Fig.~\ref{Fig:dimer}), and the other acting as the excitation acceptor ($A$, chromophore 1 in Fig.~\ref{Fig:dimer}). Let an impuslive excitation selectively excite $D$ at $t=0$. Disregarding the dynamics of thus induced optical coherences, the reduced excited-state dynamics for $t>0$ is described by
\begin{equation}
\label{Eq:rho-ee-I-Foerster-raw}
\rho_{ee}^{(I)}(t)=T\exp\left[\overrightarrow{\mathcal{W}}_p(t,0)\right]|D\rangle\langle D|,
\end{equation}
where we have dropped out the normalization constant similar to $P_e$ in Eq.~\eqref{Eq:rho-ee-spin-boson}. We are interested in the rate at which the population of $A$,
\begin{equation}
\label{Eq:P-A-t-raw}
P_A(t)=\langle A|\widetilde{U}_{DA}(t,0)\rho^{(I)}_{ee}(t)\widetilde{U}_{DA}^\dagger(t,0)|A\rangle,
\end{equation}
changes. In the last equation, a tilde over operator denotes the interaction picture with respect to the electronic Hamiltonian of the noninteracting chromophores ($\varepsilon_D|D\rangle\langle D|+\varepsilon_A|A\rangle\langle A|$), and
\begin{equation}
\widetilde{U}_{DA}(t,0)=T\exp\left[
-\frac{\im}{\hbar}\int_0^t\dif s\:\widetilde{H}_{DA}(s)
\right]
\end{equation}
with $H_{DA}=J_{DA}(|A\rangle\langle D|+|D\rangle\langle A|)$ being the $DA$ electronic coupling.

Within the F\"{o}rster theory, the population transfer from $D$ to $A$ is induced by two actions of $H_{DA}$ on opposite sides of $|D\rangle\langle D|$, while environmental DOFs are mere spectators in that process. Nevertheless, they do adapt to the change of electronic state induced by the transfer, but they alone cannot induce it if we assume (as is usual) that no environmental mode couples to both $D$ and $A$.~\cite{ChemPhys.275.319} The situation is somehow similar to the photoexcitation process, see Sec.~\ref{SSec:density-matrix}, where the excited-state sector $ee$ is reached from the ground-state sector $gg$ by applying two $H_{M-R}$ from the opposite sides of $|g\rangle\langle g|$. The phonons just adapt to the new electronic configuration, but they alone cannot bring about to the ground-to-excited state transition. Having all these things considered, it seems reasonable to attempt to replace all time-dependent operators $V_j^{(I)}(t)$ in Eq.~\eqref{Eq:rho-ee-I-Foerster-raw} by time-independent operators $V_j$ and to transform Eq.~\eqref{Eq:P-A-t-raw} by expanding $\widetilde{U}_{DA}(t,0)$ and keeping only contributions in which two $\widetilde{H}_{DA}(\tau)$ act from the opposite sides of $\rho^{(I)}_{ee}(t)$. The above-described analogy with the photoexcitation process is most conveniently exploited from the standpoint of Eq.~\eqref{Eq:final-rho-ee-formal-hyper}. The analogy then suggests that the F\"{o}rster limit be obtained by enforcing the global chronological order in the hyperoperator product acting on $|D\rangle\langle D|$, which results in the following expression for the rate of population transfer from $D$ to $A$
\begin{widetext}
\begin{equation}
\label{Eq:k-AD-F-1}
\begin{split}
k_{AD}^F(t)=\frac{2}{\hbar^2}\int_0^t\dif\tau\:\mathrm{Re}\left\{
\left\langle A\left|T\left[\widetilde{H}_{DA}(t)^C\exp\left[\overrightarrow{\mathcal{W}}_p(t,0)\right]\:^C\widetilde{H}_{DA}(\tau)\right]|D\rangle\langle D|\right|A\right\rangle
\right\}.
\end{split}
\end{equation}
We then partition the integration domain in $\overrightarrow{\mathcal{W}}_p(t,0)$ as follows [$\overrightarrow{\mathcal{F}}_p(s_2,s_1)$ denotes the hyperoperator under integral signs in Eq.~\eqref{Eq:basic-pops}]
\begin{equation}
\label{Eq:split-integration-Foerster}
\begin{split}
\overrightarrow{\mathcal{W}}_p(t,0)=
\int_0^\tau\dif s_2\int_0^{s_2}\dif s_1\:\overrightarrow{\mathcal{F}}_p(s_2,s_1)
+\int_\tau^t\dif s_2\int_\tau^{s_2}\dif s_1\:\overrightarrow{\mathcal{F}}_p(s_2,s_1)
+\int_\tau^t\dif s_2\int_0^\tau\dif s_1\:\overrightarrow{\mathcal{F}}_p(s_2,s_1).
\end{split}
\end{equation}
Let us now analyze Eq.~\eqref{Eq:k-AD-F-1} order by order in $\overrightarrow{\mathcal{W}}_p(t,0)$. This analysis bears certain resemblance to that conducted in Ref.~\onlinecite{JChemPhys.139.044102,JChemPhys.112.6719}. The approximation $V_j^{(I)}(s)\approx V_j$ is performed only after the global time-ordering prescription has been applied. Let us focus on the first-order term. The first summand on the right-hand side of Eq.~\eqref{Eq:split-integration-Foerster} describes the single-phonon assistance before the first interaction $H_{DA}$ takes place at instant $\tau$. Upon making the approximation $V_j^{(I)}(s)\approx V_j$, we conclude that the corresponding contribution is equal to zero (cf. Sec.~\ref{SSec:spin-boson}). The second summand in Eq.~\eqref{Eq:split-integration-Foerster} is effective after the first interaction $H_{DA}$, when the electronic state is that of $D/A$ coherence, $|D\rangle\langle A|$. It is then easily checked that [see also Eq.~\eqref{Eq:def-lineshape}]
\begin{equation}
\begin{split}
&T\left[\widetilde{H}_{DA}(t)^C
\int_\tau^t\dif s_2\int_\tau^{s_2}\dif s_1\:\overrightarrow{\mathcal{F}}_p(s_2,s_1)
\:^C\widetilde{H}_{DA}(\tau)
\right]|D\rangle\langle D|\approx\\ &-J_{DA}^2\:\e^{-\im(\varepsilon_D-\varepsilon_A)(t-\tau)}\left[g_D(t-\tau)+g_A^*(t-\tau)\right]|A\rangle\langle A|.
\end{split}
\end{equation}
We may anticipate that, after the resummation, this term produces the well-known factors characteristic of donor emission ($\e^{-g_D^*(t-\tau)}$) and acceptor absorption ($\e^{-g_A(t-\tau)}$). In the third summand in Eq.~\eqref{Eq:split-integration-Foerster}, one superoperator acts before, and the other after, the first interaction $H_{DA}$. This summand is expected to take into account corrections to the aforementioned donor emission factor due to the fact that donor environment has not yet adapted to the electronic excited state. In greater detail,
\begin{equation}
\begin{split}
&T\left[\widetilde{H}_{DA}(t)^C
\int_\tau^t\dif s_2\int_0^{\tau}\dif s_1\:\overrightarrow{\mathcal{F}}_p(s_2,s_1)
\:^C\widetilde{H}_{DA}(\tau)
\right]|D\rangle\langle D|\approx\\ &-J_{DA}^2\:\e^{-\im(\varepsilon_D-\varepsilon_A)(t-\tau)}\:\frac{2\im}{\hbar^2}\int_{\tau}^t\dif s_2\int_0^\tau\dif s_1\:C_D^i(s_2-s_1)|A\rangle\langle A|.
\end{split}
\end{equation}
One can convince themselves that the final result for the excitation transfer rate $k_{AD}^F(t)$ in this limit reads as
\begin{equation}
\begin{split}
k_{AD}^F(t)=\frac{2J_{DA}^2}{\hbar^2}\int_0^t\dif\tau\:
\mathrm{Re}\left\{\exp\left(\im(\varepsilon_D-\varepsilon_A)(t-\tau)-g_D^*(t-\tau)-g_A(t-\tau)+ \right. \right. \\ \left. \left. +\frac{2\im}{\hbar^2}\int_{\tau}^t\dif s_2\int_0^\tau\dif s_1\:C_D^i(s_2-s_1)\right)\right\}
\end{split}
\end{equation}
To enable a direct comparison with Eq.~(20) or Eq.~(24) of Ref.~\onlinecite{ChemPhys.275.319}, one should perform change of variables $t-\tau=\tau'$ and calculate all bath correlation functions by definition, starting from the general expression for $u_j$ [Eq.~\eqref{Eq:H-B}].
\end{widetext}

\section{Excitation by Weak Incoherent Light}
\label{Sec:weak-incoh}
Here, we study in more detail the excitation by (weak) incoherent light, for which the factorized part [Eq.~\eqref{Eq:def-G1-ij-21}] of the first-order light correlation function [Eq.~\eqref{Eq:G-1-ij-21-general}] identically vanishes. The HEOM, as formulated here, leans on the exponential decomposition of the environmental correlation function $C_j(t)$, see Eq.~\eqref{Eq:C-j-in-exp-decay}. Therefore, it may be expected that, if we can expand $G^{(1)}_{ij}(\tau_2-\tau_1)$ as a weighted sum of exponential factors, we can proceed to formulate HEOM in the usual manner. We concentrate on thermal (chaotic) light whose propagation direction and polarization are well defined. It is known that quantum and classical theory predict the same form of the first-order light correlation function for such light~\cite{Loudon-book}
\begin{equation}
\label{Eq:G-collision-broadended-CW}
G^{(1)}(\tau)=I_0\exp\left(\im\omega_c\tau-\tau/\tau_c\right).
\end{equation}
In Eq.~\eqref{Eq:G-collision-broadended-CW}, $I_0$ is the intensity, $\omega_c$ is the central frequency, while $\tau_c$ is the coherence time of the radiation. In view of the well defined polarization, we omit subscripts $i,j$ labeling Cartesian coordinates of the electric field. This form of the first-order radiation correlation function has been used to gain insight into the dynamics of open~\cite{NewJPhys.12.065044} and closed~\cite{JChemPhys.140.074104} quantum systems weakly driven by light. Here, motivated by the aforementioned exponential decomposition, we show how the following light correlation function
\begin{equation}
\label{Eq:G-exp-decomp-l}
G^{(1)}(\tau)=\sum_l I_{0,l}\exp\left(\im\omega_{c,l}\tau-\tau/\tau_{c,l}\right)
\end{equation}
can be used to recast Eq.~\eqref{Eq:final-rho-ee-formal} as HEOM.

For incoherent light, optical coherences defined in Eq.~\eqref{Eq:def-opt-coh} are exactly equal to zero. Nevertheless, the general scheme of the hierarchy is still analogous to that we outlined in the case of classical excitation, see Fig.~\ref{Fig:hierarchy}. One can introduce the following objects that act in the $eg$ sector and are thus analogous to optical coherences, cf. Eq.~\eqref{Eq:def-opt-coh},
\begin{equation}
\begin{split}
\rho_{eg,l}^{(I)}(t)&=\int_{t_0}^t\dif\tau\: U^{(I)}_\mathrm{red}(t,\tau)\frac{\im}{\hbar}\mu_{eg}^{(I)}(\tau)|g\rangle\langle g|\times\\&\times
 I_{0,l}\exp\left[\im\omega_{c,l}(t-\tau)-(t-\tau)/\tau_{c,l}\right],
\end{split}
\end{equation}
where the dipole-moment operator $\mu_{eg}$ is the projection of $\boldsymbol{\mu}_{eg}$ on the polarization direction. These optical coherence-like objects are counted by index $l$ appearing in Eq.~\eqref{Eq:G-exp-decomp-l}. In other words, each term in the exponential decomposition of the first-order radiation correlation function adds a new layer to the HEOM for "optical coherences", which reads as
\begin{equation}
\label{Eq:heom-eg-incoh}
 \begin{split}
 \partial_t\sigma_{eg,l,\mathbf{n}}(t)=&-\frac{\im}{\hbar}\left[H_M,\sigma_{eg,l,\mathbf{n}}(t)\right]\\&+
 \left(\im\omega_{c,l}-\tau_{c,l}^{-1}\right)\sigma_{eg,l,\mathbf{n}}(t)\\&-
 \left(\sum_j\sum_m n_{j,m}\mu_{j,m}\right)\sigma_{eg,l,\mathbf{n}}(t)\\&+
 \delta_{\mathbf{n},\mathbf{0}}\frac{\im}{\hbar}I_{0,l}\mu_{eg}\\
 &+\im\sum_j\sum_m V_j\sigma_{eg,l,\mathbf{n}_{j,m}^+}(t)\\&+\im\sum_{j}\sum_{m}n_{j,m}\frac{c_{j,m}}{\hbar^2}V_j\sigma_{eg,l,\mathbf{n}_{j,m}^-}(t).
 \end{split}
\end{equation}
Nevertheless, the HEOM for singly excited-state populations and intraband coherences does not feature any additional layers stemming from the decomposition in Eq.~\eqref{Eq:G-exp-decomp-l} and it reads as
\begin{equation}
\label{Eq:heom-ee-incoh}
 \begin{split}
  &\partial_t\sigma_{ee,\mathbf{n}}(t)=-\frac{\im}{\hbar}\left[H_M,\sigma_{ee,\mathbf{n}}(t)\right]\\
  &-\left(\sum_j\sum_m n_{j,m}\mu_{j,m}\right)\sigma_{ee,\mathbf{n}}(t)\\
  &+\frac{\im}{\hbar}\mu_{eg}\left(\sum_l\sigma^\dagger_{eg,l,\mathbf{n}}(t)\right)-
  \frac{\im}{\hbar}\left(\sum_l\sigma_{eg,l,\mathbf{n}}(t)\right)\mu_{ge}\\
  &+\im\sum_j\sum_m\left[V_j,\sigma_{ee,\mathbf{n}_{j,m}^+}(t)\right]\\
  &+\im\sum_j\sum_m n_{j,m}\frac{c_{j,m}}{\hbar^2}\:V_j\sigma_{ee,\mathbf{n}_{j,m}^-}(t)\\
  &-\im\sum_j\sum_m n_{j,m}\frac{c_{j,m}^*}{\hbar^2}\sigma_{ee,\mathbf{n}_{j,m}^-}(t)V_j.
 \end{split}
\end{equation}

Our results embodied in Eqs.~\eqref{Eq:heom-eg-incoh} and~\eqref{Eq:heom-ee-incoh} are significant because they provide a viable route towards a description of excitonic dynamics triggered by thermal light. This description consistently combines both specific temporal and statistical properties of the radiation and a nonperturbative treatment of the excitation--environment coupling. It is well known that, in principle, the only sensible representation of thermal light is statistical, in terms of a set of all possible realizations. The recently suggested representation of natural incoherent light as an ensemble of transform-limited pulses~\cite{JChemPhys.144.044103} would suggest that the dynamics it induces be computed by propagating HEOM embodied in Eqs.~\eqref{Eq:y-x-heom-semiclassical} and~\eqref{Eq:n-barx-x-heom-semiclassical} for individual ensemble realizations and then averaging over them. This would present a formidable task, since propagating Eqs.~\eqref{Eq:y-x-heom-semiclassical} and~\eqref{Eq:n-barx-x-heom-semiclassical} for one ensemble member is already numerically expensive. Equations~\eqref{Eq:heom-eg-incoh} and~\eqref{Eq:heom-ee-incoh} demonstrate how such complications can be circumvented within the second-order treatment of the interaction with light. The fact that the exponential decomposition of the first-order radiation correlation function does not affect the complexity of HEOM in the excited-state sector is numerically advantageous. Namely, propagating HEOM for "optical coherences" [Eq.~\eqref{Eq:heom-eg-incoh}] is significantly less numerically demanding than propagating HEOM in the excited-state sector [Eq.~\eqref{Eq:heom-ee-incoh}]. Since the dynamics of "optical coherences" in our second-order treatment is not affected by the dynamics in the excited-state sector, we conclude that the overall numerical complexity of the problem as we formulate it is not significantly greater than in the case of pulsed photoexcitation [Eqs.~\eqref{Eq:y-x-heom-semiclassical} and~\eqref{Eq:n-barx-x-heom-semiclassical}].

The proposed theory is valid in the limit of weak light--matter interaction. To provide a more quantitative criterion of this weakness, we recall that the maximum normal surface solar irradiance at sea level on a clear day is $\mathcal{I}_\mathrm{max}\approx 1~\mathrm{kW}/\mathrm{m}^2$.~\cite{wald:hal-01676634} The electric field amplitude corresponding to this irradiance can be estimated by $E_0=\sqrt{\frac{2\mathcal{I}_\mathrm{max}}{c\varepsilon_0}}$ ($c$ is the speed of light, while $\varepsilon_0$ is the vacuum permittivity) and we obtain $E_0\approx 870$~V/m. Keeping in mind that the magnitude of the transition dipole moment of the bacteriochlorophyll molecule is $d_{eg}\approx 6$~D,~\cite{Photosynthetic-excitons-book} we can estimate the magnitude of the interaction energy of electronic excitations and radiation by $E_0 d_{eg}\sim 10^{-3}~\mathrm{cm}^{-1}$. We see that this interaction energy is orders of magnitude smaller than the energies characteristic for excitonic couplings, exciton-environment interactions, and static disorder in transition energies ($\sim$10--100~$\mathrm{cm}^{-1}$).~\cite{RevModPhys.90.035003} Keeping in mind that the lunar irradiance or the solar irradiance in habitats of some photosynthetic bacteria are even smaller than the maximum solar irradiance upon which the above estimates were based, we conclude that the weak-light assumption is well satisfied in various photosynthetically relevant situations. The same conclusion may be reached by estimating the rate of solar photons incident on a photosynthetic complex. To that end, we start from the fact that the normal surface solar irradiance of the photosynthetically available radiation ($400-700\:\mathrm{nm}$) of the solar spectrum is $\mathcal{I}_\mathrm{PAR}\approx 540\:\mathrm{W}/\mathrm{m}^2$.~\cite{wald:hal-01676634} In typical photosynthetic complexes, bacteriochlorophyll molecules most strongly absorb at wavelengths around $\lambda\approx 700-800\:\mathrm{nm}$.~\cite{Photosynthetic-excitons-book} Taking that the typical linear dimension of a photosynthetic complex is $a\sim 10$~\AA,~\cite{RevModPhys.90.035003} we may estimate that the number of solar photons incident on a complex per unit time is $\frac{dN}{dt}=\mathcal{I}_\mathrm{PAR}\frac{\lambda a^2}{2\pi\hbar c}\simeq 2000\:\mathrm{s}^{-1}$, which agrees well with the estimate provided in Ref.~\onlinecite{JChemPhys.152.154101}. The actual photon absorption rate, which also depends on the effective absorption cross section and the degree of radiation attenuation due to the specific habitat conditions, may be even smaller. The corresponding temporal scale is thus orders of magnitude longer than time scales typical for EET, excitation recombination or extraction, which corroborates the plausibility of our weak-light assumption.

In the following, we compare our Eqs.~\eqref{Eq:heom-eg-incoh} and~\eqref{Eq:heom-ee-incoh} with the existing descriptions of  photoexcitation by incoherent light.~\cite{JPhysChemLett.3.3136,JPhysB.51.054002,JChemTheorComput.7.2166} We concentrate on the light correlation function in Eq.~\eqref{Eq:G-collision-broadended-CW}. The aforementioned approaches exploit the fact that the coherence time of natural Sunlight $\tau_c\sim 1\:\mathrm{fs}$~\cite{ProcPhysSoc.80.1273,IlNuovoCimento.28.401} is at least an order of magnitude shorter than the time scales typical for electronic couplings and nuclear reorganization processes (which assume values $\sim 10-100\:\mathrm{cm}^{-1}$).~\cite{RevModPhys.90.035003} We may then argue that we can disregard the nonequilibrium environmental dynamics taking place between the two interactions with the radiation, which was crucial to correctly describe excitonic dynamics induced by a pulsed photoexcitation, see the discussion accompanying Fig.~\ref{Fig:primitive-propagators}(c) and Figs.~\ref{Fig:heom-redfield}(a1)--\ref{Fig:heom-redfield}(d2). In other words, we may assume that both interactions with the radiation occur essentially at the same instant, which means that Eq.~\eqref{Eq:G-collision-broadended-CW} should be replaced by
\begin{equation}
\label{Eq:G-WNM}
 G^{(1)}(\tau)=2I_0\tau_c\:\delta(\tau).
\end{equation}
This is the so-called white-noise model (WNM) of the radiation.~\cite{Olsina:2014}
In this case, one has to propagate only the HEOM for excited-state populations and interband coherences and Eq.~\eqref{Eq:heom-ee-incoh} should be replaced by
\begin{equation}
 \label{Eq:heom-ee-incoh-wnm}
 \begin{split}
  \partial_t\sigma_{ee,\mathbf{n}}(t)=&-\frac{\im}{\hbar}\left[H_M,\sigma_{ee,\mathbf{n}}(t)\right]\\
  &-\left(\sum_j\sum_m n_{j,m}\mu_{j,m}\right)\sigma_{ee,\mathbf{n}}(t)\\
  &+\delta_{\mathbf{n},\mathbf{0}}\:\frac{2I_0\tau_c}{\hbar^2}\mu_{eg}|g\rangle\langle g|\mu_{ge}\\
  &+\im\sum_j\sum_m\left[V_j,\sigma_{ee,\mathbf{n}_{j,m}^+}(t)\right]\\
  &+\im\sum_j\sum_m n_{j,m}\frac{c_{j,m}}{\hbar^2}\:V_j\sigma_{ee,\mathbf{n}_{j,m}^-}(t)\\
  &-\im\sum_j\sum_m n_{j,m}\frac{c_{j,m}^*}{\hbar^2}\sigma_{ee,\mathbf{n}_{j,m}^-}(t)V_j.
 \end{split}
\end{equation}

\begin{figure}
 \includegraphics{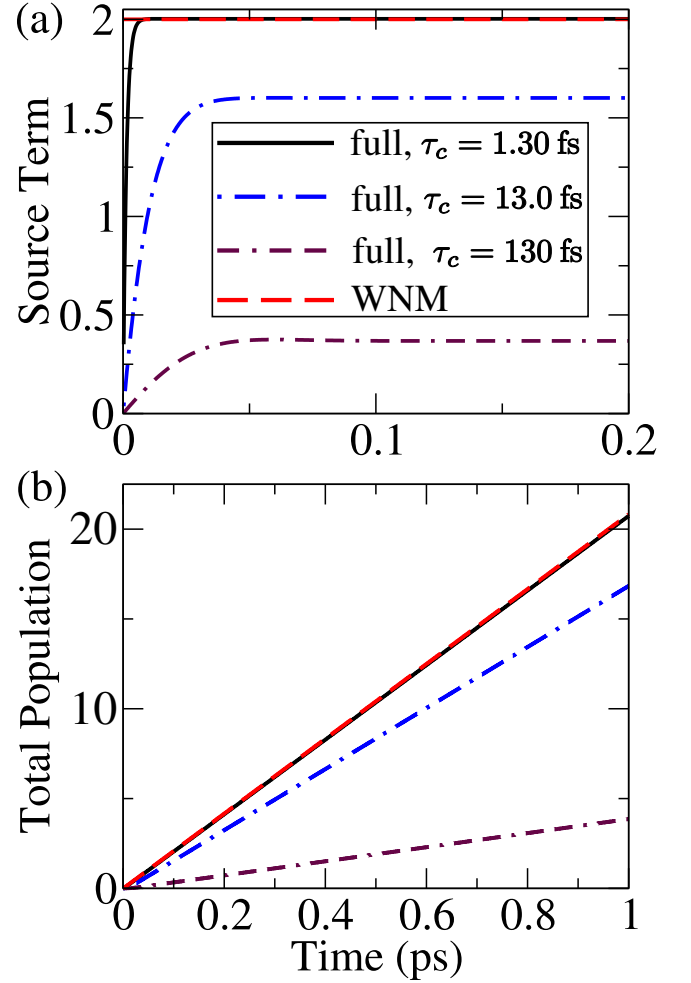}
 \caption{(Color online) (a) Source term [in units of $\gamma\tau_c I_0d_{eg}^2/(\hbar\gamma)^2$] in the equation for the total excited-state population of the model dimer (see Fig.~\ref{Fig:dimer}) when the dynamics is governed by
 Eq.~\eqref{Eq:heom-ee-incoh-wnm} (WNM, dashed line) and when the dynamics is governed by Eqs.~\eqref{Eq:heom-eg-incoh} and~\eqref{Eq:heom-ee-incoh} for $\tau_c=1.30\:\mathrm{fs}$ (solid line), $\tau_c=13.0\:\mathrm{fs}$ (dash-dotted line), and $\tau_c=130\:\mathrm{fs}$ (double dash-dotted line). (b) Total excited-state population [in units of $\gamma\tau_c I_0d_{eg}^2/(\hbar\gamma)^2$] after a sudden turn-on of incoherent light with $\tau_c=1.30\:\mathrm{fs}$ (solid line), $\tau_c=13.0\:\mathrm{fs}$ (dash-dotted line), $\tau_c=130\:\mathrm{fs}$ (double dash-dotted line) and $\tau_c\to 0$ (WNM, dashed line).}
 \label{Fig:full-vs-wnm}
\end{figure}

In Figs.~\ref{Fig:full-vs-wnm}(a) and~\ref{Fig:full-vs-wnm}(b) we confront the dynamics of the dimer described in Fig.~\ref{Fig:dimer} that is triggered by suddenly turned on incoherent light and governed by Eqs.~\eqref{Eq:heom-eg-incoh} and~\eqref{Eq:heom-ee-incoh} (label "full") and Eq.~\eqref{Eq:heom-ee-incoh-wnm} (label "WNM") for different values of light coherence time $\tau_c$.
Figure~\ref{Fig:full-vs-wnm}(a) presents the dimensionless source term for the total excited-state population for $G^{(1)}(\tau)$ given in Eq.~\eqref{Eq:G-collision-broadended-CW} (curves labeled as ``full'')
\begin{equation}
 S_\mathrm{full}(t)=\frac{\im}{\hbar}\mathrm{Tr}_M\left\{\mu_{eg}\sigma^\dagger_{eg,\mathbf{0}}(t)-
  \sigma_{eg,\mathbf{0}}(t)\mu_{ge}\right\}
\end{equation}
and for $G^{(1)}(\tau)$ given in Eq.~\eqref{Eq:G-WNM} (the curve labeled as ``WNM'')
\begin{equation}
 S_\mathrm{WNM}=\frac{2I_0\tau_c}{\hbar^2}\mathrm{Tr}_M\left\{\mu_{eg}|g\rangle\langle g|\mu_{ge}\right\}.
\end{equation}
Instead of presenting the data in absolute units, the dimensionless source term in Fig.~\ref{Fig:full-vs-wnm}(a) and the total population in Fig.~\ref{Fig:full-vs-wnm}(b) are given in units of $\gamma\tau_c I_0d_{eg}^2/(\hbar\gamma)^2$. The reason for choosing this unit is our second-order treatment of exciton--light interaction, which ensures that the excited-state populations and intraband coherences scale linearly in the light intensity $I_0$ [Eqs.~\eqref{Eq:G-collision-broadended-CW},~\eqref{Eq:G-exp-decomp-l} and~\eqref{Eq:G-WNM}] and quadratically in the transition dipole moment $d_{eg}$ (see Fig.~\ref{Fig:dimer}). On the formal side, one can convince themselves that $\gamma\tau_c I_0d_{eg}^2/(\hbar\gamma)^2$ is the appropriate unit for our purposes by analyzing Eqs.~\eqref{Eq:heom-eg-incoh},~\eqref{Eq:heom-ee-incoh} and~\eqref{Eq:heom-ee-incoh-wnm}. The presence of the factor $\gamma\tau_c$ in the unit enables us to compare on the same plot the data for different light coherence times. The value of this unit is estimated by recalling that the energy scale of the exciton--light interaction is $E_0d_{eg}\sim 10^{-3}\:\mathrm{cm}^{-1}$ and that we use $\gamma^{-1}=100\:\mathrm{fs}$ (see the caption of Fig.~\ref{Fig:dimer}). Therefore, for $\tau_c=1.3\:\mathrm{fs}$, $\gamma\tau_c I_0d_{eg}^2/(\hbar\gamma)^2\simeq 4\times 10^{-12}$, for $\tau_c=13\:\mathrm{fs}$, $\gamma\tau_c I_0d_{eg}^2/(\hbar\gamma)^2\simeq 4\times 10^{-11}$, while for $\tau_c=130\:\mathrm{fs}$, $\gamma\tau_c I_0d_{eg}^2/(\hbar\gamma)^2\simeq 4\times 10^{-10}$. The value of $\tau_c$ determines the time scale on which the source term for the total excited-state population $S_\mathrm{full}(t)$ reaches a constant value upon a sudden turn-on of incoherent radiation, see Fig.~\ref{Fig:full-vs-wnm}(a). For the shortest $\tau_c$ examined, the source term $S_\mathrm{full}(t)$ saturates within the initial $\sim 10\:\mathrm{fs}$ of the dynamics and the value it reaches excellently agrees with the source term $S_\mathrm{WNM}$ predicted by the WNM of the radiation. This is also reflected in Fig.~\ref{Fig:full-vs-wnm}(b), in which the total exciton populations predicted by the two models display a perfect agreement for the shortest $\tau_c$. As $\tau_c$ is increased, so that it becomes comparable to time scales of nuclear reorganization processes, the agreement between the results predicted by Eqs.~\eqref{Eq:heom-eg-incoh} and~\eqref{Eq:heom-ee-incoh} and Eq.~\eqref{Eq:heom-ee-incoh-wnm} deteriorates, see Figs.~\ref{Fig:full-vs-wnm}(a) and~\ref{Fig:full-vs-wnm}(b), because the WNM cannot capture the nonequilibrium bath dynamics between the two interactions with the radiation. The larger is the coherence time $\tau_c$, the more pronounced are the deviations of the exact dynamics from the WNM results.

\begin{widetext}
%\onecolumngrid

\begin{figure}
 \includegraphics[scale=0.8]{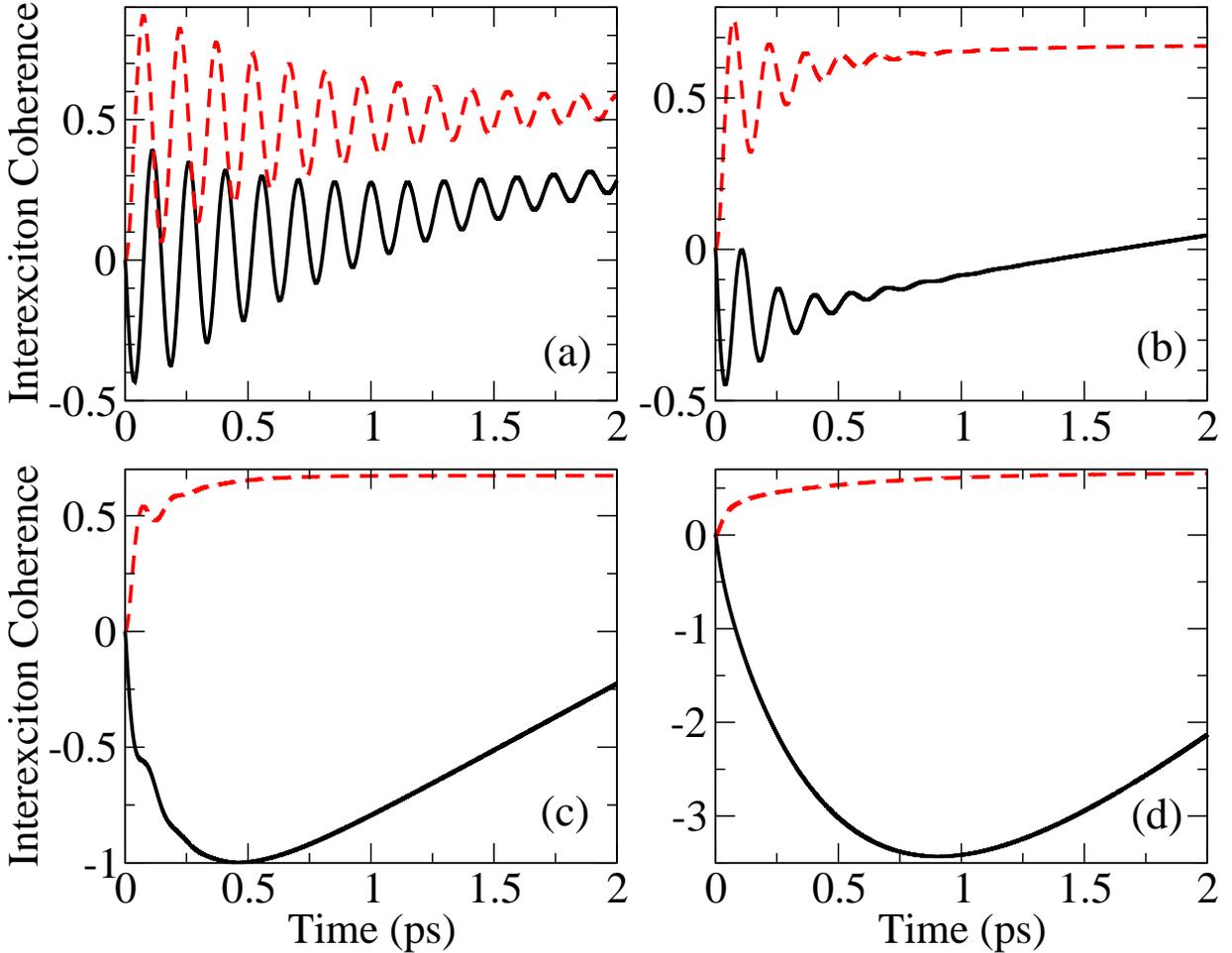}
 \caption{(Color online) Dynamics of the real (solid lines) and imaginary (dashed lines) parts of the interexciton coherence [in units of $\gamma\tau_cI_0d_{eg}^2/(\hbar\gamma)^2$] in the model dimer (see Fig.~\ref{Fig:dimer}) for different values of the reorganization energy: (a) $\lambda=2\:\mathrm{cm}^{-1}$, (b) $\lambda=20\:\mathrm{cm}^{-1}$, (c) $\lambda=100\:\mathrm{cm}^{-1}$, and (d) $\lambda=500\:\mathrm{cm}^{-1}$. Light coherence time is $\tau_c=1.3\:\mathrm{fs}$.}
 \label{Fig:interexc-thermal}
\end{figure}
%\twocolumngrid
\end{widetext}

We now turn our attention to the dynamics of interexciton coherences, which is displayed in Figs.~\ref{Fig:interexc-thermal}(a)--\ref{Fig:interexc-thermal}(d) for different values of the reorganization energy $\lambda$.

The initial (sub-picosecond) oscillatory dynamics of the interexciton coherence that is clearly observed for lower values of $\lambda$, see Figs.~\ref{Fig:interexc-thermal}(a) and~\ref{Fig:interexc-thermal}(b), is directly related to the oscillations displayed by the populations in the site basis upon an ultrafast excitation, see Figs.~\ref{Fig:heom-redfield}(a2) and~\ref{Fig:heom-redfield}(b2).~\cite{JChemPhys.130.234110} As the reorganization energy is increased, the oscillatory features gradually disappear, cf. Figs.~\ref{Fig:heom-redfield}(b1)--\ref{Fig:heom-redfield}(d2), and certain steady behavior of the interexciton coherence, similar to a steady increase in the total exciton population observed in Fig.~\ref{Fig:full-vs-wnm}, sets in. In Figs.~\ref{Fig:interexc-thermal}(b)--\ref{Fig:interexc-thermal}(d), we see that the imaginary part of the interexction coherence saturates in $\sim 1\:\mathrm{ps}$ after the excitation start. On the other hand, the real part of the interexciton coherence in Fig.~\ref{Fig:interexc-thermal}(b) exhibits a steady increase for $t\gtrsim 1.25\:\mathrm{ps}$, while the corresponding start of the steady increased is shifted to $\sim 0.75\:\mathrm{ps}$ and $\sim 1.5\:\mathrm{ps}$ in Figs.~\ref{Fig:interexc-thermal}(c) and~\ref{Fig:interexc-thermal}(d), respectively. The time scale on which the steady increase of the interexciton coherence sets in is intimately related to the time scale on which the populations in the site basis [see Figs.~\ref{Fig:heom-redfield}(b2)--\ref{Fig:heom-redfield}(d2)] reach their limiting values following a very short photoexcitation. By virtue of basis transformation,~\cite{JChemPhys.130.234110} the latter is closely connected to the time scale of the dephasing of the interexciton coherence generated by a very short photoexcitation. Therefore, the oscillatory features under incoherent illumination originate from the sudden turn-on of the excitation at $t=0$~\cite{JChemPhys.145.244313,JPhysChemLett.9.2946} and they disappear on the time scale on which the interexciton coherence dephases after a short photoexcitation.~\cite{Olsina:2014} We note, in passing, that the magnitude of the interexciton coherence becomes much larger than populations, which is also in line with previous studies.~\cite{JChemPhys.145.244313,JPhysChemLett.9.2946,Olsina:2014,JChemPhys.144.044103} The light-induced coherences observed in Figs.~\ref{Fig:interexc-thermal}(a) and~\ref{Fig:interexc-thermal}(b) are not expected to be directly relevant to excitation harvesting under natural conditions, which proceeds via nonequilibrium steady states.~\cite{JPhysChemLett.9.2946} Such states arise from a combination of the steady increase in populations and coherences [see Fig.~\ref{Fig:full-vs-wnm}(b) and Figs.~\ref{Fig:interexc-thermal}(a)--(d)] due to continuous generation and the steady excitation decay by trapping at the reaction center and recombination. What may be relevant for the efficiency of light harvesting driven by incoherent light are the relations among the RDM elements that establish themselves on time scales typical for excitation trapping and recombination, which are generally much longer than the time scales covered in this study. Further development of this idea is the topic of the accompanying paper.~\cite{ness-vj-tm}

This section explicitly deals with incoherent light whose first-order correlation function is of the form given in Eqs.~\eqref{Eq:G-collision-broadended-CW} and~\eqref{Eq:G-exp-decomp-l}. In the literature,~\cite{JChemTheorComput.7.2166,JPhysChemLett.3.3136,JPhysB.51.054002} it is common to formulate equations similar to Eq.~\eqref{Eq:heom-ee-incoh-wnm}, which tacitly lean on the WNM in which the source term describing the generation of state $|e\rangle$ (from the ground state) by incoherent Sunlight is given in terms of the number of photons of energy $\varepsilon_e$ at the temperature of the Sun's photosphere ($\sim 6000\:\mathrm{K}$). In Sec.~SIV of the Supplementary Material, we demonstrate how our description reduces to the above-described quantum-optical limit by virtue of the Weisskopf--Wigner approximation,~\cite{Scully-Zubairy-book} which has to be performed in the exciton basis.~\cite{Breuer-Petruccione-book} In essence, exploiting the weakness of the excitation--light coupling, the quantum-optical approaches tacitly assume that the effects of this coupling can be simply added to the dynamics in the absence of radiation in form of Markovian corrections that do not feature any modifications due to the presence of the environment.~\cite{JChemTheorComput.7.2166,JPhysChemLett.3.3136,JPhysB.51.054002} While such approaches do include a radiative recombination term, our approach does not contain such a term. As demonstrated in greater detail in Sec.~SIV of the Supplementary Material, the reason for the absence of the radiative recombination term in Eqs.~\eqref{Eq:heom-eg-incoh} and~\eqref{Eq:heom-ee-incoh} or in Eq.~\eqref{Eq:heom-ee-incoh-wnm} is the fact that our dynamics starts from the initially unexcited system and that it consistently keeps track of interactions with light up to the second order. To describe nonequilibrium stationary states under incoherent light, our theoretical framework has to be augmented by additional drain terms that should take into account excitation recombination and possibly some other mechanisms by which the excitations may be lost (e.g., trapping at the reaction center). In our opinion, this feature of our formalism does not render it less suitable for steady-state calculations under incoherent light, see the accompanying paper.~\cite{ness-vj-tm} However, a fully self-consistent approach to obtain the steady-state under incoherent light is still out of our reach because it asks for a solution to the fundamental problem of the nonadditivity of the excitation--light and excitation--environment coupling.~\cite{PhysRevLett.123.093601}

\section{Conclusion}
\label{Sec:conclusion}
We have conducted a detailed theoretical investigation of the dynamics of electronic excitations in molecular aggregates induced by weak radiation of arbitrary properties. Starting from initially unexcited aggregate, our approach combines a perturbative treatment of the coupling to radiation with an exact treatment of the excitation--environment coupling in a manner that is manifestly compatible with the spectroscopic view of the photoexcitation. We express the reduced excited-state dynamics entirely in terms of the first-order radiation correlation function and the reduced evolution superoperator,
for which we provide an exact expression within the Frenkel exciton model.
The changes that the state of electronic excitations undergoes due to the photoexcitation and the interaction with the environment can be conveniently represented diagrammatically, in terms of elementary processes assisted by single quanta of environmental excitations.
The fact that our general expression for the excited-state dynamics explicitly keeps track of the instants at which the two interactions with light occur means that the corresponding differential equation is time-nonlocal. Within the exponential decomposition scheme, we outline how this temporal nonlocality can be circumvented by setting up a suitable HEOM scheme that explicitly takes into account the photoexcitation step. Such developments, however, turn out to heavily depend on radiation properties. 

In the case of excitation by transform-limited pulses, when the radiation correlation function factorizes into product of (classical) electric fields at two interaction instants, we relate HEOM arising from our results to the HEOM obtained by considering the (semiclassical) light--matter coupling as a part of the aggregate Hamiltonian. The insights from nonlinear spectroscopy and semiconductor optics analyzed using the DCT scheme turn out to be crucial in establishing that relationship. Namely, the order in which the light--matter coupling is taken into account determines the subspace of the excitation Fock space on which the photoinduced dynamics should be formulated, and \emph{vice versa}. We demonstrate that the second-order response to light should be formulated on the subspace containing at most one excitation. The second-order results [Eqs.~\eqref{Eq:y-x-heom-semiclassical} and~\eqref{Eq:n-barx-x-heom-semiclassical}] presented in Sec.~\ref{Sec:heom-arb-e-t} may be extended to treat laser-induced nonlinear effects of an arbitrary order and yet retain the numerically exact treatment of the interaction of photoinduced excitations with the environment. We believe that this result may be of relevance for future treatments of laser-induced coherent EET.

We analyze in detail the dynamics triggered by an impulsive excitation of a single chromophore, where we provide analytical results for environmental reorganization dynamics, which is encoded in ADMs. We further identify the approximations under which our general result reduces to the widely employed Redfield theory with photoexcitation and nonequilibrium F\"{o}rster theory. Our comparison between the dynamics predicted by HEOM and Redfield theory with photoexcitation further corroborates the advantages of our approach, which exactly describes light-induced environmental reorganization processes and fully takes into account the nonequilibrium evolution of the bath between the two interactions with light.

In the case of excitation by thermal light, we compare our approach to the widely used hybrid Born--Markov--HEOM approach, which treats the light--matter coupling within quantum-optical approximations. Since we employ the exponential decomposition scheme, the formulation of HEOM relies on an exponential decomposition of the radiation correlation function. We find that the HEOM thus obtained is not significantly more numerically expensive than the HEOM as it is usually formulated. This is because additional layers stemming from the decomposition of radiation correlation function exist only in its optical-coherence-like part, which may be solved completely independently from its excited-state part. This paves the way towards viable simulations of the dynamics triggered by natural incoherent light that respect both the specific properties of the radiation and the need for a nonperturbative treatment of excitation--environment coupling.

We believe that, despite its unfavorable numerical cost, the approach outlined here can be useful in further investigations of light--induced dynamics in both photosynthetic light-harvesting aggregates and OPVs. In particular, in both types of systems the relation of the insights gained in ultrafast spectroscopies to the actual operation under natural Sunlight illumination has provoked long-standing debates. Our method promises to bridge these two standpoints and establish a new more suitable common viewpoint on energy conversion in these systems. Further work on bridging the two standpoints is under way in our respective research groups.

\section*{Supplementary Material}
See supplementary material for: (a) a more detailed discussion of the second-order response to excitation by coherent light; (b) a detailed derivation of the exact evolution superoperator; (c) derivation of the Redfield equation comprising photoexcitation; (d) a discussion of the quantum-optical limit in the case of excitation by incoherent light.

\begin{acknowledgments}
We gratefully acknowledge the support by Charles University Research Centre program No. UNCE/SCI/010. T.M. acknowledges the funding by the Czech Science Foundation (GA\v CR) grant No. 18-18022S. The final stages of this work were supported by the Institute of Physics Belgrade, through the grant by the Ministry of Education, Science, and Technological Development of the Republic of Serbia.
\end{acknowledgments}

\section*{Author's Contributions}
V.J. and T.M. designed the research. V.J. performed all the analytical and numerical work under the supervision of T.M. V.J. prepared the initial version of the manuscript. Both authors contributed to the submitted version of the manuscript.

\section*{Data Availability Statement}
The data that support the findings of this study are available from the corresponding authors upon reasonable request.

\bibliography{apssamp}% Produces the bibliography via BibTeX.
\end{document}

\title{Supplementary Material for:\\
Exact description of excitonic dynamics in molecular aggregates weakly driven by light}

\author{Veljko Jankovi\'{c}}
\email{veljko.jankovic@ipb.ac.rs}
\affiliation{
 Faculty of Mathematics and Physics, Charles University, Ke Karlovu 5, 121 16 Prague 2, Czech Republic
}
\affiliation{Scientific Computing Laboratory, Center for the Study of Complex Systems, Institute of Physics Belgrade, University of Belgrade, Pregrevica 118, 11080 Belgrade, Serbia}
\author{Tom\'{a}\v{s} Man\v{c}al}%
 \email{mancal@karlov.mff.cuni.cz}
\affiliation{
 Faculty of Mathematics and Physics, Charles University, Ke Karlovu 5, 121 16 Prague 2, Czech Republic
}

\maketitle

\section{Second-Order Response within Semiclassical Treatment of Light--Matter Interaction}
\label{Sec:2nd-order-response}
In this section, we explain in greater detail why, when we study the second-order response, we can safely limit ourselves to the subspace containing at most one excitation.

The full density matrix $W(t)$ describing the combined system of aggregate excitations and environment can be expanded in powers $n$ of the exciting field
\begin{equation}
W(t)=\sum_{n=0}^{+\infty} W^{(n)}(t),\text{ where }W^{(n)}(t)\propto\mathcal{E}^n.
\end{equation}
According to the expansion theorem of Ref.~\onlinecite{PhysRevB.53.7244}, the $n$th order contribution $W^{(n)}(t)$ may be expanded in terms of states containing a definite number of excitations as follows
\begin{equation}
\label{Eq:expansion-theorem}
W^{(n)}(t)=\sum_{n'=0}^n\sum_{\kappa'\kappa}|n',\kappa',t\rangle\langle n-n',\kappa,t|\:\rho_B(n,n',\kappa,\kappa',t).
\end{equation}
Here, $|n',\kappa',t\rangle$ is a number state containing $n'$ excitations characterized by quantum numbers that are collectively denoted as $\kappa'$, while its explicit time dependence stems from the time dependence of the exciting field. In that sense, state $|n',\kappa',t\rangle$ is not normalized, but rather scales as $\mathcal{E}^{n'}$. The expansion coefficients $\rho_B(n,n',\kappa,\kappa',t)$ are purely environmental operators.

Using the expansion theorem, it can be shown that the second-order response is fully formulated in terms of the following generating functions~\cite{RevModPhys.70.145}
\begin{equation}
    Y_j^{\alpha\beta}=\left\langle B_j\hat{F}^{\alpha\beta}\right\rangle,
\end{equation}
\begin{equation}
\label{Eq:def-N-ij-alphabeta}
    N_{ij}^{\alpha\beta}=\left\langle B_i^\dagger B_j\hat{F}^{\alpha\beta}\right\rangle,
\end{equation}
where
\begin{equation}
    \hat{F}^{\alpha\beta}=\exp\left(\sum_{j\xi}\alpha_{j\xi}b_{j\xi}^\dagger\right)\exp\left(\sum_{j\xi}\beta_{j\xi}b_{j\xi}\right).
\end{equation}
$Y_j^{\alpha\beta}$ is the generating function for (environment-assisted) optical coherences, while $N_{ij}^{\alpha\beta}$ is the generating function for (environment-assisted) singly excited-state populations and intraband coherences. A more elaborate analysis shows that (environment-assisted) biexcitonic amplitudes, $\left\langle B_i B_j \hat{F}^{\alpha\beta}\right\rangle$, which also scale as $\mathcal{O}(\mathcal{E}^2)$, do not contribute to the second-order response, so that they are omitted from further discussions.

Let us now recall that the excitation annihilation operator $B_j$ may be expanded in terms of number states so that it manifestly connects subspaces accommodating different numbers of excitons~\cite{Haug-Jauho-book}
\begin{equation}
B_j=|0\rangle\langle 1,j|+\sum_{\substack{k\\k\neq j}}|1,k\rangle\langle 2,kj|+\dots
\end{equation}
Then,
\begin{equation}
\begin{split}
Y_j^{\alpha\beta}&=\sum_{n=0}^{+\infty}\sum_{n'=0}^n\sum_{\kappa'\kappa}\mathrm{Tr}_M\left\{B_j|n',\kappa',t\rangle\langle n-n',\kappa,t|\right\}\:\mathrm{Tr}_B\left\{\hat{F}^{\alpha\beta}\rho_B(n,n',\kappa,\kappa',t)\right\}\\
&=\sum_{n=0}^{+\infty}\sum_{n'=0}^n\sum_{\kappa'\kappa}\langle 1,j|n',\kappa',t\rangle\langle n-n',\kappa,t|0\rangle\:\mathrm{Tr}_B\left\{\hat{F}^{\alpha\beta}\rho_B(n,n',\kappa,\kappa',t)\right\}+\\
&+\sum_{n=0}^{+\infty}\sum_{n'=0}^n\sum_{\kappa'\kappa}\sum_{\substack{k\\k\neq j}}\langle 2,kj|n',\kappa',t\rangle\langle n-n',\kappa,t|1,k\rangle\:\mathrm{Tr}_B\left\{\hat{F}^{\alpha\beta}\rho_B(n,n',\kappa,\kappa',t)\right\}+\dots
\end{split}
\end{equation}
We see that the summand containing $\langle 1,j|n',\kappa',t\rangle\langle n-n',\kappa,t|0\rangle$ is nonzero iff $n=n'=1$, i.e., in the first order in the exciting field. The summand containing $\langle 2,kj|n',\kappa',t\rangle\langle n-n',\kappa,t|1,k\rangle$ is nonzero iff $n'=2$ and $n=3$, i.e., in the third order in the optical field. Therefore, the expansion of the excitation annihilation operator in number states actually reflects the contributions to $Y_j^{\alpha\beta}$ that are linear, cubic, etc. in the exciting field, as predicted by the central theorem of the DCT scheme.~\cite{RevModPhys.70.145} The fact that the environment has no impact on scaling relations is apparent in our discussion. Up to the second order in the exciting field, contributions to $B_j$ that involve a state containing more than one particle do not contribute to the response.

In a similar vein,
\begin{equation}
\label{Eq:B-dagger-B-in-Hilbert}
B_i^\dagger B_j=|1,i\rangle\langle 1,j|+\sum_{\substack{k\\k\neq i,k\neq j}}|2,ik\rangle\langle 2,jk|+\dots
\end{equation}
Combining Eqs.~\eqref{Eq:expansion-theorem},~\eqref{Eq:def-N-ij-alphabeta}, and~\eqref{Eq:B-dagger-B-in-Hilbert}, we conclude that all contributions to $B_i^\dagger B_j$ involving states that accommodate more than a single excitation are at least of the fourth order in the exciting field, and, therefore, do not contribute to the second-order response.

If we are to study the second-order response, we can limit our description to subspaces that contain up to one excitation. On the other hand, as described in the main text, once we formulate the model Hamiltonian in the subspace containing at most one excitation, we can consistently formulate the dynamics only up to the second order in the exciting field. 

\vspace*{\fill}
\pagebreak

\section{Derivation of the Exact Evolution Superoperator}
Here, we present the derivation of the exact evolution superoperator for a weakly driven excitonic aggregate.

In the interaction picture, the total DM $W^{(I)}(t)$ of the combined system comprising the molecular aggregate, its environment, and the radiation field, evolves according to
\begin{equation}
\label{Eq:von-Neumann-total-interaction}
 \partial_t W^{(I)}(t)=-\frac{\im}{\hbar}\left[H_{M-B}^{(I)}(t)+H_{M-R}^{(I)}(t),W^{(I)}(t)\right],
\end{equation}
with the factorized initial condition $W^{(I)}(t_0)\equiv W(t_0)$ given in the main text. Here, we define the interaction picture with respect to the non-interacting Hamiltonian
\begin{equation}
\label{Eq:non-inter-H}
 H_0=H_M+H_B+H_R,
\end{equation}
so that for any operator $O$ in the Schr\"{o}dinger picture, the corresponding operator in the interaction picture with respect to $H_0$ reads as
\begin{subequations}
\label{Eq:def-interaction-picture-all}
\begin{eqnarray}
O^{(I)}(t)=U_0^\dagger(t,t_0)O U_0(t,t_0),\label{Eq:def-interaction-picture}
\\
U_0(t,t_0)=\exp\left[-\frac{\im}{\hbar}H_0(t-t_0)\right].\label{Eq:def-U0}
\end{eqnarray}
\end{subequations}
The formal solution to Eq.~\eqref{Eq:von-Neumann-total-interaction} is
\begin{widetext}
\begin{equation}
\label{Eq:formal_W_I}
 \begin{split}
  W^{(I)}(t)=\sum_{n=0}^{+\infty}\left(-\frac{\im}{\hbar}\right)^n
  \int_{t_0}^t\dif\tau_n\dots\int_{t_0}^{\tau_2}\dif\tau_1
  \left[H_{M-B}^{(I)}(\tau_n)+H_{M-R}^{(I)}(\tau_n),\dots,\left[H_{M-B}^{(I)}(\tau_1)+H_{M-R}^{(I)}(\tau_1),W(t_0)\right]\dots\right].
 \end{split}
\end{equation}
\end{widetext}
Let us now focus on the case of weak interaction with the radiation by keeping in Eq.~\eqref{Eq:formal_W_I} only contributions that contain no more than two interaction Hamiltonians $H_{M-R}$. At the same time, as discussed in Sec.~\ref{Sec:2nd-order-response}, this means that we can safely reduce our description to the subspace containing at most one excitation and consequently make the replacements $B_j\to|g\rangle\langle j|,B_j^\dagger\to|j\rangle\langle g|$ in the model Hamiltonian. The specific form of the initial condition, as well as the fact that any nontrivial dynamics is ultimately induced by $H_{M-R}$, enable us to separately treat different electronic sectors ($gg,eg,ge,$ and $ee$) of the total DM. By electronic sectors $gg$, $eg$, and $ee$ of the total DM, we understand here its parts that, after appropriate reductions, contain information on the ground-state population, optical coherences, and excited-state populations and intraband coherences, respectively. After a straightforward analysis, we obtain the following results for the $eg$ sector
\begin{equation}
\label{Eq:W-eg-formal-a}
 W^{(I)}_{eg}(t)=-\frac{\im}{\hbar}\int_{t_0}^t\dif\tau\:U_{M-B}^{(I)}(t,\tau)H^{(I)}_{M-R}(\tau)W(t_0),
\end{equation}
for the $gg$ sector
\begin{equation}
\label{Eq:W-gg}
\begin{split}
 W^{(I)}_{gg}(t)&=W(t_0)
 -\frac{\im}{\hbar}\int_{t_0}^t\dif\tau\: H^{(I)}_{M-R}(\tau)W^{(I)}_{eg}(\tau)\\
 &+\frac{\im}{\hbar}\int_{t_0}^t\dif\tau\: W^{(I)\dagger}_{eg}(\tau)H^{(I)}_{M-R}(\tau),
 \end{split}
\end{equation}
and for the $ee$ sector
\begin{equation}
\label{Eq:W-ee-feynman-paths}
\begin{split}
 W^{(I)}_{ee}(t)&=\int_{t_0}^t\dif\tau_2\int_{t_0}^{\tau_2}\dif\tau_1\:
 U^{(I)}_{M-B}(t,\tau_2)U^{(I)}_{M-B}(\tau_2,\tau_1)
 \left(\frac{1}{\hbar^2}H^{(I)}_{M-R}(\tau_1)W(t_0)H^{(I)}_{M-R}(\tau_2)\right)U^{(I)\dagger}_{M-B}(t,\tau_2)\\
 &+\int_{t_0}^t\dif\tau_2\int_{t_0}^{\tau_2}\dif\tau_1\:
 U^{(I)}_{M-B}(t,\tau_2)\left(\frac{1}{\hbar^2}H^{(I)}_{M-R}(\tau_2)W(t_0)H^{(I)}_{M-R}(\tau_1)\right)
 U^{(I)\dagger}_{M-B}(\tau_2,\tau_1)U^{(I)\dagger}_{M-B}(t,\tau_2).
 \end{split}
\end{equation}
In Eqs.~\eqref{Eq:W-eg-formal-a} and~\eqref{Eq:W-ee-feynman-paths},
\begin{equation}
\label{Eq:def-U-M-B-21}
U_{M-B}^{(I)}(s_2,s_1)=T\exp\left[-\frac{\im}{\hbar}\int_{s_1}^{s_2}\dif s\:H^{(I)}_{M-B}(s)\right],
\end{equation}
where $T$ denotes the chronological time ordering. The result embodied in Eq.~\eqref{Eq:W-ee-feynman-paths} can be interpreted in terms of double-sided Feynman diagrams.~\cite{Mukamel-book} The two summands on the right-hand side of Eq.~\eqref{Eq:W-ee-feynman-paths} represent the two Liouville pathways from $|g\rangle\langle g|$ to $|e\rangle\langle e|$, which differ by the time order of the interactions with the bra and ket. These two summands are complex conjugates of one another, so that $W^{(I)}_{ee}(t)$ is a Hermitian operator.~\cite{ChemPhys.404.103} Therefore, in further discussions, it will be enough to perform manipulations with the first summand only.

Let us note that the total number of excitations, which is given by the (total) trace of $W(t)$ [or $W^{(I)}(t)$], is conserved. This is most easily proven by rewriting Eqs.~\eqref{Eq:W-ee-feynman-paths} and~\eqref{Eq:W-gg} as differential equations.

The RDM containing excited-state populations and intraband coherences is obtained by performing partial traces with respect to the radiation and the thermal bath
\begin{equation}
 \rho_{ee}^{(I)}(t)=\mathrm{Tr}_B\left\{\mathrm{Tr}_R\left\{W_{ee}^{(I)}(t)\right\}\right\}.
\end{equation}
Let us concentrate on reducing the first term on the right-hand side of Eq.~\eqref{Eq:W-ee-feynman-paths}. The partial trace over radiation is computed straightforwardly, since radiation operators enter Eq.~\eqref{Eq:W-ee-feynman-paths} only through the two $H_{M-R}$ terms. In more detail,
\begin{equation}
\frac{1}{\hbar^2}\mathrm{Tr}_R\left\{H^{(I)}_{M-R}(\tau_1)W(t_0)H^{(I)}_{M-R}(\tau_2)\right\}=A^{(I)}(\tau_2,\tau_1)\rho_B^g,
\end{equation}
where the purely electronic operator $A^{(I)}(\tau_2,\tau_1)$ is defined as
\begin{equation}
\label{Eq:def-A-I-21}
% \begin{split}
 A^{(I)}(\tau_2,\tau_1)=
 \frac{1}{\hbar^2}\sum_{i,j}G^{(1)}_{ij}(\tau_2,\tau_1)\left\{\boldsymbol{\mu}_{eg}^{(I)}(\tau_1)\right\}_j|g\rangle\langle g|\left\{\boldsymbol{\mu}_{ge}^{(I)}(\tau_2)\right\}_i.
% \end{split}
\end{equation}
In Eq.~\eqref{Eq:def-A-I-21}, the sums over $i$ and $j$ are performed over Cartesian components of the electric field, whereas $G^{(1)}_{ij}(\tau_2,\tau_1)$ is defined in the main text. Therefore, in the limit of weak aggregate--radiation interaction, the radiation enters the reduced dynamics of excited-state populations and intraband coherences only via its first-order correlation function, as has already been demonstrated.~\cite{NewJPhys.12.065044} Integrating over the bath degrees of freedom provides us with the second ingredient governing the excitonic dynamics, which is the reduced evolution superoperator explicitly containing the two interaction instants $\tau_1$ and $\tau_2$ with the radiation, as well as the observation instant $t$. Having taken partial trace over radiation, we now expand each phonon propagator [Eq.~\eqref{Eq:def-U-M-B-21}] entering Eq.~\eqref{Eq:W-ee-feynman-paths} in powers of $H_{M-B}$. Since the averaging is performed over the canonical density matrix of the bath, Wick's theorem ensures that only contributions containing an even number of $H_{M-B}$ operators should be evaluated and that the final result is entirely expressed in terms of the two-point (noninteracting) bath correlation function $C_j(t)$ defined in the main text.~\cite{Mahan-book,Fetter-Walecka-book}

To simplify further discussion, we concentrate on the following operator
%(interaction instants $\tau_1$ and $\tau_2$ with radiation are fixed)
\begin{equation}
w^{(I)}_{ee,1}(t,\tau_2,\tau_1)=U^{(I)}_{M-B}(t,\tau_2)U^{(I)}_{M-B}(\tau_2,\tau_1)A^{(I)}(\tau_2,\tau_1)\rho_B^g U^{(I)\dagger}_{M-B}(t,\tau_2),
\end{equation}
whose partial trace over bath has to be computed. The operator $w^{(I)}_{ee,1}(t,\tau_2,\tau_1)$ describes the state of electronic excitations and the environment %, the reduction over radiation has already been performed)
at time $t$ when we assume that the first interaction with the radiation takes place at $\tau_1$ from the left, while the second interaction occurs at $\tau_2$ from the right. Upon expanding $U^{(I)}_{M-B}$ in powers of $H^{(I)}_{M-B}$, we obtain
\begin{equation}
\label{Eq:pathway-1-general-1}
\begin{split}
w^{(I)}_{ee,1}(t,\tau_2,\tau_1)=&\sum_{n=0}^{+\infty}\left(-\frac{\im}{\hbar}\right)^n\frac{1}{n!}\sum_{m=0}^{+\infty}\left(-\frac{\im}{\hbar}\right)^m\frac{1}{m!}\int_{\tau_2}^t\dif q_n\dots\int_{\tau_2}^t\dif q_1\int_{\tau_1}^{\tau_2}\dif s_m\dots\int_{\tau_1}^{\tau_2}\dif s_1\\&T\left[H^{(I)}_{M-B}(q_n)^\times\dots H^{(I)}_{M-B}(q_1)^\times\right]\:T\left[H^{(I)}_{M-B}(s_m)^C\dots H^{(I)}_{M-B}(s_1)^C\right]A^{(I)}(\tau_2,\tau_1)\rho_B^g,
\end{split}
\end{equation}
where hyperoperators with superscritps $\times$ and $C$ have the same meaning as in the main text. The index $n$ controls the order of expansion of the combination $U^{(I)}_{M-B}(t,\tau_2)\dots U^{(I)\dagger}_{M-B}(t,\tau_2)$, and the instants at which individual excitation--environment interactions take place are denoted as $q_1,\dots,q_n$. Similarly, index $m$ controls the order of expansion of $U^{(I)}_{M-B}(\tau_2,\tau_1)$, and the respective instants are $s_1,\dots,s_m$. We now use the explicit form of the excitation--environment coupling
\begin{equation}
H^{(I)}_{M-B}(s_1)=\sum_{l_1} V_{l_1}^{(I)}(s_1)u_{l_1}^{(I)}(s_1),
\end{equation}
where $V_{l_1}^{(I)}(s_1)$ is a purely electronic operator, while $u_{l_1}^{(I)}(s_1)$ is a purely environmental operator. Let us focus in the following on the operator under sums and integrals in Eq.~\eqref{Eq:pathway-1-general-1}. Since $A^{(I)}(\tau_2,\tau_1)$ is a purely electronic, while $\rho_B^g$ is a purely environmental operator, we can write
\begin{equation}
\label{Eq:pathway-1-general-2}
\begin{split}
&T\left[H^{(I)}_{M-B}(q_n)^\times\dots H^{(I)}_{M-B}(q_1)^\times\right]\:T\left[H^{(I)}_{M-B}(s_m)^C\dots H^{(I)}_{M-B}(s_1)^C\right]A^{(I)}(\tau_2,\tau_1)\rho_B^g=\\
&\sum_{j_n\dots j_1}\sum_{l_m\dots l_1}
T\left[\left(V_{j_n}^{(I)}(q_n)u_{j_n}^{(I)}(q_n)\right)^\times\dots\left(V_{j_1}^{(I)}(q_1)u_{j_1}^{(I)}(q_1)\right)^\times\right]\times\\
&\times\underbrace{T\left[V_{l_m}^{(I)}(s_m)^C\dots V_{l_1}^{(I)}(s_1)^C\right]A^{(I)}(\tau_2,\tau_1)}_{\text{purely electronic}}\:
\underbrace{T\left[u_{l_m}^{(I)}(s_m)^C\dots u_{l_1}^{(I)}(s_1)^C\right]\rho_B^g}_{\text{purely environmental}},
\end{split}
\end{equation}
where, in each term of the sum, the hyperoperators from the outermost layer act on an operator that is factorized into a purely electronic and purely environmental part.
At this point, the following identity is useful
\begin{equation}
(Vu)^\times O_V O_u=\frac{1}{2}\left(V^\times u^\circ+V^\circ u^\times\right)O_V O_u,
\end{equation}
where $O_V$ and $O_u$ are arbitrary operators such that $V$ can act on $O_V$ only, while $u$ can act on $O_u$ only. The last identity enables us to rewrite Eq.~\eqref{Eq:pathway-1-general-2} as follows
\begin{equation}
\begin{split}
&T\left[H^{(I)}_{M-B}(q_n)^\times\dots H^{(I)}_{M-B}(q_1)^\times\right]\:T\left[H^{(I)}_{M-B}(s_m)^C\dots H^{(I)}_{M-B}(s_1)^C\right]A^{(I)}(\tau_2,\tau_1)\rho_B^g=\\
&\frac{1}{2^n}\sum_{j_n\dots j_1}\sum_{l_m\dots l_1}
T\left[\left(V_{j_n}^{(I)}(q_n)^\times u_{j_n}^{(I)}(q_n)^\circ+V_{j_n}^{(I)}(q_n)^\circ u_{j_n}^{(I)}(q_n)^\times\right)\dots\left(V_{j_1}^{(I)}(q_1)^\times u_{j_1}^{(I)}(q_1)^\circ+V_{j_1}^{(I)}(q_1)^\circ u_{j_1}^{(I)}(q_1)^\times\right)\right]\times\\
&\times\underbrace{T\left[V_{l_m}^{(I)}(s_m)^C\dots V_{l_1}^{(I)}(s_1)^C\right]A^{(I)}(\tau_2,\tau_1)}_{\text{purely electronic}}\:
\underbrace{T\left[u_{l_m}^{(I)}(s_m)^C\dots u_{l_1}^{(I)}(s_1)^C\right]\rho_B^g}_{\text{purely environmental}}.
\end{split}
\end{equation}
Each term under multiple sums over $j$s and $l$s has $2^n$ summands and, in each of them, the hyperoperators from the outermost layer can be written as
$$V_{j_n}^{(I)}(q_n)^{\bar{\sigma}_n}u_{j_n}^{(I)}(q_n)^{\sigma_n}\dots V_{j_1}^{(I)}(q_1)^{\bar{\sigma}_1}u_{j_1}^{(I)}(q_1)^{\sigma_1},$$
where $\sigma_i\in\{\times,\circ\}$, $\bar{\times}=\circ$, and \emph{vice versa}.
Moreover, since all instants $q_n,\dots,q_1$ are certainly later than all instants $s_m,\dots,s_1$, we can merge the two time-ordering signs (one ordering $q$s and the other ordering $s$s) into a single time-ordering sign. Having all these transformations performed, we are left with
\begin{equation}
\begin{split}
&T\left[H^{(I)}_{M-B}(q_n)^\times\dots H^{(I)}_{M-B}(q_1)^\times\right]\:T\left[H^{(I)}_{M-B}(s_m)^C\dots H^{(I)}_{M-B}(s_1)^C\right]A^{(I)}(\tau_2,\tau_1)\rho_B^g=\\
&\frac{1}{2^n}\sum_{j_n\dots j_1}\sum_{l_m\dots l_1}\sum_{\sigma_n\dots\sigma_1}
T\left[V_{j_n}^{(I)}(q_n)^{\bar{\sigma}_n}\dots V_{j_1}^{(I)}(q_1)^{\bar{\sigma}_1}
V_{l_m}^{(I)}(s_m)^C\dots V_{l_1}^{(I)}(s_1)^C\right]A^{(I)}(\tau_2,\tau_1)\times\\
&\times T\left[u_{j_n}^{(I)}(q_n)^{\sigma_n}\dots u_{j_1}^{(I)}(q_1)^{\sigma_1}u_{l_m}^{(I)}(s_m)^C\dots u_{l_1}^{(I)}(s_1)^C\right]\rho_B^g.
\end{split}
\end{equation}
We have thus reduced the problem to the computation of the following trace over the environment
\begin{equation}
\label{Eq:partial-bath-hyperoperator-1}
\mathrm{Tr}_B\left\{T\left[u_{j_n}^{(I)}(q_n)^{\sigma_n}\dots u_{j_1}^{(I)}(q_1)^{\sigma_1}u_{l_m}^{(I)}(s_m)^C\dots u_{l_1}^{(I)}(s_1)^C\right]\rho_B^g\right\}.
\end{equation}
It is clear that this trace may give a nontrivial result only when $n+m$ is even, $n+m=2k$, where $k=1,2\dots$. Let us now start from the simplest case $n+m=2$, in which we face the following possibilities:
\begin{enumerate}
\item $n=2$ and $m=0$: environmental assistance is characteristic for the electronic subsystem in the excited-state manifold
\begin{equation}
\label{Eq:basic-block-populations}
\begin{split}
\mathrm{Tr}_B\left\{T\left[u_{j_2}^{(I)}(q_2)^{\sigma_2}u_{j_1}^{(I)}(q_1)^{\sigma_1}\right]\rho_B^g\right\}&=
\theta(q_2-q_1)\delta_{j_2j_1}\cdot 2^2\left(\delta_{\sigma_2,\circ}\delta_{\sigma_1,\circ}C_{j_1}^r(q_2-q_1)+\delta_{\sigma_2,\circ}\delta_{\sigma_1,\times}\im C_{j_1}^i(q_2-q_1)\right)+\\
&+\theta(q_1-q_2)\delta_{j_2j_1}\cdot 2^2\left(
\delta_{\sigma_2,\circ}\delta_{\sigma_1,\circ}C_{j_1}^r(q_1-q_2)+
\delta_{\sigma_2,\times}\delta_{\sigma_1,\circ}\im C_{j_1}^i(q_1-q_2).
\right)
\end{split}
\end{equation}
The corresponding contribution to the partial trace over environment of Eq.~\eqref{Eq:pathway-1-general-1} then reads as
\begin{equation}
\begin{split}
\left[\mathrm{Tr}_B w^{(I)}_{ee,1}(t,\tau_2,\tau_1)\right]_{n=2,m=0}&=-\frac{1}{\hbar^2}\sum_j\int_{\tau_2}^t\dif q_2\int_{\tau_2}^{q_2}\dif q_1\:\times\\&\times V_j^{(I)}(q_2)^\times\left(C_j^r(q_2-q_1)V_j^{(I)}(q_1)^\times+\im C_j^i(q_2-q_1)V_j^{(I)}(q_1)^\circ\right)A^{(I)}(\tau_2,\tau_1)\\
&\equiv\overrightarrow{\mathcal{W}}_p(t,\tau_2)A^{(I)}(\tau_2,\tau_1),
\end{split}
\end{equation}
which is the familiar form that has been obtained when the propagation is considered only within the excited-state manifold.~\cite{JChemPhys.130.234111}
\item $n=0$ and $m=2$: environmental assistance is characteristic for the electronic subsystem in the state of optical coherence
\begin{equation}
\label{Eq:basic-block-coherence}
\begin{split}
\mathrm{Tr}_B\left\{
T\left[u_{l_2}^{(I)}(s_2)^C u_{l_1}^{(I)}(s_1)^C\right]\rho_B^g
\right\}&=\theta(s_2-s_1)\delta_{l_2 l_1}C_{l_1}(s_2-s_1)+\theta(s_1-s_2)\delta_{l_2 l_1}C_{l_1}(s_1-s_2)\\&\equiv\delta_{l_1 l_2}C_{l_1}(|s_2-s_1|).
\end{split}
\end{equation}
The corresponding contribution to the partial trace over environment of Eq.~\eqref{Eq:pathway-1-general-1} then reads as
\begin{equation}
\begin{split}
\left[\mathrm{Tr}_B w^{(I)}_{ee,1}(t,\tau_2,\tau_1)\right]_{n=0,m=2}&=-\frac{1}{\hbar^2}\sum_l\int_{\tau_1}^{\tau_2}\dif s_2\int_{\tau_1}^{s_2}\dif s_1\:V_l^{(I)}(s_2)^C\:C_l(s_2-s_1)V_l^{(I)}(s_1)^C\:A^{(I)}(\tau_2,\tau_1)\\
&\equiv\overrightarrow{\mathcal{W}}_c(\tau_2,\tau_1)A^{(I)}(\tau_2,\tau_1).
\end{split}
\end{equation}
\item $n=1$ and $m=1$: environmental assistance straddles over periods in which electronic subsystem is in different states: it starts when the electronic subsystem is in the state of optical coherence, and ends when the electronic subsystem is in the excited-state manifold; here, we obtain the elementary contribution to the so-called straddled evolution
\begin{equation}
\label{Eq:basic-block-straddled}
\mathrm{Tr}_B\left\{
T\left[u_{j_1}^{(I)}(q_1)^{\sigma_1}u_{l_1}^{(I)}(s_1)^C\right]\rho_B^g
\right\}=\delta_{\sigma_1,\circ}\delta_{j_1,l_1}\cdot 2C_{j_1}(q_1-s_1).
\end{equation}
Note that $T$ sign in the last equation can safely be omitted, since we are sure that $q_1\geq s_1$. The corresponding contribution to the partial trace over environment of Eq.~\eqref{Eq:pathway-1-general-1} then reads as
\begin{equation}
\begin{split}
\left[\mathrm{Tr}_B w^{(I)}_{ee,1}(t,\tau_2,\tau_1)\right]_{n=1,m=1}&=-\frac{1}{\hbar^2}\sum_j\int_{\tau_2}^{t}\dif q_1\int_{\tau_1}^{\tau_2}\dif s_1\:V_j^{(I)}(q_1)^\times\:C_j(q_1-s_1)V_j^{(I)}(s_1)^C\:A^{(I)}(\tau_2,\tau_1)\\&\equiv\overrightarrow{\mathcal{W}}_{c-p}(t,\tau_2,\tau_1)A^{(I)}(\tau_2,\tau_1).
\end{split}
\end{equation}
\end{enumerate}
The above analysis conducted in the lowest order of the perturbation expansion gives us basic building blocks from which higher-order contributions are constructed. This is possible by virtue of the Wick's theorem
\begin{equation}
\label{Eq:Wick-th-operator}
\mathrm{Tr}_B\left\{T\left[
u_{j_{2k}}^{(I)}(q_{2k})\dots u_{j_1}^{(I)}(q_1)
\right]\rho_B^g
\right\}=\sum_\text{a.p.p.}\prod_{a,b}\mathrm{Tr}_B\left\{
T\left[u_{j_b}^{(I)}(q_b)u_{j_a}^{(I)}(q_a)\right]\rho_B^g
\right\},
\end{equation}
which expresses the (equilibrium) expectation value of the product of $2k$ nuclear displacement operators $u$ as a sum over all possible pairings (a.p.p.) of products of $k$ (equilibrium) expectation values of two nuclear displacement operators. A similar identity also holds on the hyperoperator level, which is relevant to our discussion [see Eq.~\eqref{Eq:partial-bath-hyperoperator-1}]. Namely, since all hyperoperators $u^\pi$, where $\pi\in\left\{\times,\circ,C\right\}$ are linear in nuclear displacement $u$, we can use Eq.~\eqref{Eq:Wick-th-operator} to establish the following identity
\begin{equation}
\label{Eq:Wick-th-hyperoperator}
\mathrm{Tr}_B\left\{T\left[
u_{j_{2k}}^{(I)}(q_{2k})^{\pi_{2k}}\dots u_{j_1}^{(I)}(q_1)^{\pi_1}
\right]\rho_B^g
\right\}=\sum_\text{a.p.p.}\prod_{a,b}\mathrm{Tr}_B\left\{
T\left[u_{j_b}^{(I)}(q_b)^{\pi_b}u_{j_a}^{(I)}(q_a)^{\pi_a}\right]\rho_B^g
\right\}.
\end{equation}

Equation~\eqref{Eq:Wick-th-hyperoperator} provides us with a general recipe to compute the partial trace in Eq.~\eqref{Eq:partial-bath-hyperoperator-1} and, eventually, evaluate the contribution to Eq.~\eqref{Eq:pathway-1-general-1} in arbitrary perturbation order defined by values of $n$ and $m$. The form of Eq.~\eqref{Eq:Wick-th-hyperoperator} suggests that all the contributions are indeed expressed in terms of the three elementary (lowest-order) contributions that we have evaluated. However, we should still convince ourselves that all these contributions can be resummed into the form of a time-ordered exponential that presented in the main body of the manuscript. 

Since this demonstration is rather formal and not particularly insightful, we only make its sketch for $n+m=2k$, $k\geq 1$. There, we have $2k+1$ different possibilities for $n$ and $m$. In the sketch, we make use of the so-called polynomial expansion, which states that, for mutually commuting entities $x_c,x_p,x_{c-p}$ we have
\begin{equation}
(x_c+x_p+x_{c-p})^k=\sum_{\substack{l_c+l_p+l_{c-p}=k\\l_c,l_p,l_{c-p}\geq 0}}\frac{k!}{l_c!l_p!l_{c-p}!}\:x_c^{l_c}x_p^{l_p}x_{c-p}^{l_{c-p}}.
\end{equation}
The commutativity in our case is ensured by the presence of the global time-ordering sign.
\begin{enumerate}
\item $n=2k$, $m=0$\\
The application of Wick's theorem [Eq.~\eqref{Eq:Wick-th-hyperoperator}] produces $(2k-1)!!$ terms, and it turns out that all of them are the same. In essence, this follows from the fact that all integrals over $q_{2k},\dots,q_1$ are over the same interval $[\tau_2,t]$ and that $T$ sign enables us to permute at will the hyperoperators it affects. All the contributions contain environmental assistance in the form characteristic for the electronic subsystem in the excited-state manifold. Even though the Heaviside functions in Eq.~\eqref{Eq:basic-block-populations} order the integration variables in pairs, they do not enforce the global order, so that $T$ sign cannot be removed.

The factor $2^{-2k}$ (appearing in front of sums by $j$s and $l$s) is cancelled by the factor $(2^2)^k$ that emerges from $k$ factors of the type given in Eq.~\eqref{Eq:basic-block-populations}. Then, the total prefactor is
$$\frac{1}{(2k)!}\cdot (2k-1)!!=\frac{1}{k!}\cdot\frac{1}{2^k},$$
where $(k!)^{-1}$ is indicative of the $k$-th order in the expansion of an exponential, whereas $(1/2)^k$ compensates for two equivalent time orderings in Eq.~\eqref{Eq:basic-block-populations}. In Fig.~\ref{fig:f1} we give the corresponding diagrams for $k=2$.
\begin{figure}[htbp]
    \centering
    \includegraphics[scale=1.25]{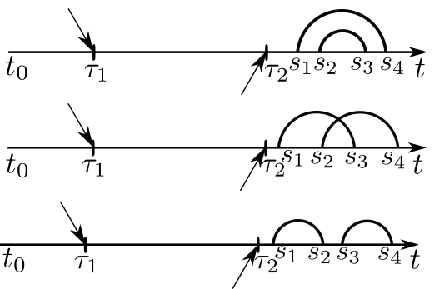}
    \caption{Diagrams representing different contributions in the fourth-order in $H_{M-B}$ when $n=4$ and $m=0$. Their sum can be understood as the square of the primitive diagram in Fig. 1(b) of the main body of the manuscript. For simplicity, here, we omit information on the state of the electronic subsystem ($|g\rangle$ or $|e\rangle$).}
    \label{fig:f1}
\end{figure}
\item $n=2k-1$, $m=1$\\
Here, we have one environmental assistance that is mixed (straddled), as in Eq.~\eqref{Eq:basic-block-straddled}, and $(k-1)$ assistances that are characteristic for the electronic system in the excited-state manifold, as in Eq.~\eqref{Eq:basic-block-populations}. All the terms produced by the application of Wick's theorem again turn out to be the same. The factor $2^{-(2k-1)}$ (appearing in front of sums by $j$s and $l$s) is cancelled by the factor $(2^2)^{k-1}\cdot 2$ which stems from $(k-1)$ factors similar to that in Eq.~\eqref{Eq:basic-block-populations} and one factor similar to that in Eq.~\eqref{Eq:basic-block-straddled}. The overall prefactor
$$\frac{1}{(2k-1)!}\cdot\frac{1}{1!}\cdot(2k-1)!!=\frac{1}{2^{k-1}}\cdot k\cdot\frac{1}{k!}$$
is then combined with the appropriate term as follows. The factor $(1/2)^{k-1}$ compensates for two equivalent time orderings in each of $(k-1)$ factors analogous to Eq.~\eqref{Eq:basic-block-populations} (the straddled assistance, by its definition, features a definite time ordering), factor $k$ reflects the fact that there are $k$ equivalent ways of choosing the straddled building block (this is the polynomial coefficient $k!/(l_c!l_p!l_{c-p}!)$ for $l_c=0,l_p=k-1,l_{c-p}=1$), while $(k!)^{-1}$ is again indicative of the $k$-th order in the expansion of an exponential. In Fig.~\ref{fig:f2} we give the corresponding diagrams for $k=2$.
\begin{figure}[htbp]
    \centering
    \includegraphics[scale=1.25]{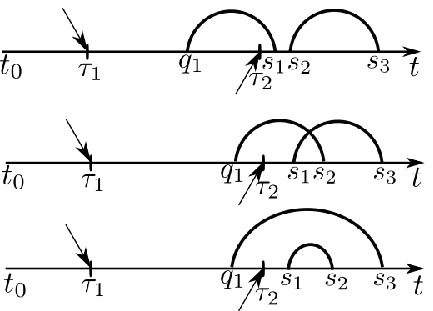}
    \caption{Diagrams representing different contributions in the fourth-order in $H_{M-B}$ when $n=3$ and $m=1$. Their sum can be understood as twice the product of the primitive diagrams in Figs. 1(b) and 1(c) of the main body of the manuscript. For simplicity, here, we omit information on the state of the electronic subsystem ($|g\rangle$ or $|e\rangle$).}
    \label{fig:f2}
\end{figure}
\item $n=2k-2$, $m=2$\\
Here, we may have
\begin{itemize}
    \item[(a)] $(k-1)$ population-like assistances and 1 coherence-like assistance; the number of such terms, which are all mutually identical, is $$(2(k-1)-1)!!=(2k-3)!!;$$
    \item[(b)] $(k-2)$ population-like assistances and 2 straddled assistances; the number of such terms, which are all mutually identical, is $$(2k-2)(2k-3)(2(k-2)-1)!!=(2k-2)(2k-3)!!.$$
\end{itemize}
Of course, the total number of terms produced by the application of Wick's theorem is $(2k-1)!!$. Let us briefly comment on the way how the prefactors combine.
\begin{itemize}
    \item[(a)] the prefactor $2^{-(2k-2)}$ (appearing in front of sums over $j$s and $l$s) is cancelled by the factor $(2^2)^{k-1}$ stemming from $(k-1)$ terms like that in Eq.~\eqref{Eq:basic-block-populations}; the overall prefactor
    $$\frac{1}{(2k-2)!}\cdot\frac{1}{2!}\cdot(2k-3)!!=\frac{1}{2^k}\cdot k\cdot\frac{1}{k!}$$
    is combined with the appropriate terms as follows. The factor $(1/2)^k$ compensates for two equivalent time orderings in each of $(k-1)$ factors analogous to Eq.~\eqref{Eq:basic-block-populations} and the remaining factor analogous to Eq.~\eqref{Eq:basic-block-coherence}; factor $k$ is again the polynomial coefficient $k!/(l_c!l_p!l_{c-p}!)$ for $l_c=1,l_p=k-1,l_{c-p}=0$;
    \item[(b)] the prefactor $2^{-(2k-2)}$ (appearing in front of sums over $j$s and $l$s) is cancelled by the product $(2^2)^{k-2}\cdot 2^2$ originating from $(k-2)$ factors like that in Eq.~\eqref{Eq:basic-block-populations} and 2 factors like that in Eq.~\eqref{Eq:basic-block-straddled}; in the overall prefactor
    $$\frac{1}{(2k-2)!}\cdot\frac{1}{2!}\cdot(2k-2)(2k-3)!!=\frac{1}{2^{k-2}}\binom{k}{2}\frac{1}{k!},$$
    $(1/2)^{k-2}$ compensates for two equivalent time orderings [Eq.~\eqref{Eq:basic-block-populations}], while $\binom{k}{2}$ is the polynomial coefficient $k!/(l_c!l_p!l_{c-p}!)$ for $l_c=0,l_p=k-2,l_{c-p}=2$.  
 \end{itemize}
In Figs.~\ref{fig:f3}(a) and~\ref{fig:f3}(b) we give the corresponding diagrams for $k=2$. 
\begin{figure}[htbp]
    \centering
    \includegraphics[scale=1.25]{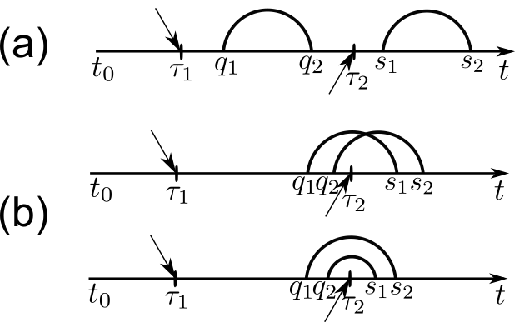}
    \caption{Diagrams representing different contributions in the fourth-order in $H_{M-B}$ when $n=2$ and $m=2$. The diagram in (a) corresponds to the contribution analyzed under point (a) of the discussion and can be understood as twice the product of primitive diagrams in Figs. 1(a) and 1(b) of the main body of the manuscript. The two diagrams in (b) correspond to the contributions analyzed under point (b) of the discussion and can be understood as the square of the primitive diagram in Fig. 1(c) of the main body of the manuscript. For simplicity, here, we omit information on the state of the electronic subsystem ($|g\rangle$ or $|e\rangle$).}
    \label{fig:f3}
\end{figure} 
\end{enumerate}
In a similar manner, one can analyze the remaining terms in order $2k$. The main result of such an analysis is the presence of the factor $1/k!$, as well as the corresponding polynomial factor $k!/(l_c!l_p!l_{c-p}!)$, in each of these terms. Therefore, the resummation produces the time-ordered exponential that is presented in the main text.

Finally, let us present the expression for the reduced evolution superoperator $\overleftarrow{\mathcal{U}}^{(I)}_\mathrm{red}(t,\tau_2,\tau_1)$ that acts on the left ($\overline{T}$ denotes antichronological time order)
\begin{subequations}
\label{Eq:U-left-t-2-1-all}
\begin{eqnarray}
\overleftarrow{\mathcal{U}}_\mathrm{red}^{(I)}(t,\tau_2,\tau_1)=\overline{T}\exp\left[\overleftarrow{\mathcal{W}}_c(\tau_2,\tau_1)+\overleftarrow{\mathcal{W}}_p(t,\tau_2)+\overleftarrow{\mathcal{W}}_{c-p}(t,\tau_2,\tau_1)\right],\label{Eq:U-left-t-2-1}
\\
\overleftarrow{\mathcal{W}}_c(\tau_2,\tau_1)=-\frac{1}{\hbar^2}\sum_j\int_{\tau_1}^{\tau_2}\dif s_2\int_{\tau_1}^{s_2}\dif s_1\:^C V_j^{(I)}(s_1)\:C_j(s_1-s_2)\:^C V_j^{(I)}(s_2),\label{Eq:basic-coh-left}
\\
\overleftarrow{\mathcal{W}}_p(t,\tau_2)=-\frac{1}{\hbar^2}\sum_j\int_{\tau_2}^{t}\dif s_2\int_{\tau_2}^{s_2}\dif s_1\left(C_j^r(s_1-s_2)\:^\times V_j^{(I)}(s_1)+\im\:C_j^i(s_1-s_2)\:^\circ V_j^{(I)}(s_1)\right)\:^\times V_j^{(I)}(s_2),\label{Eq:basic-pops-left}
\\
\overleftarrow{\mathcal{W}}_{c-p}(t,\tau_2,\tau_1)=-\frac{1}{\hbar^2}\sum_j\int_{\tau_2}^t\dif s_2\int_{\tau_1}^{\tau_2}\dif s_1\:^C V_j^{(I)}(s_1)^\times\:C_j(s_1-s_2)\:^\times V_j^{(I)}(s_2).\label{Eq:basic-pops-cohs-left}
\end{eqnarray}
\end{subequations}
In Eqs.~\eqref{Eq:U-left-t-2-1-all}, we define hyperoperators $^{\times/\circ/C}V$ acting on the left (which is suggested by the position of the superscript) by the following equalities valid for any operator $O$
\begin{subequations}
\label{Eq:define-hyper}
\begin{eqnarray}
O\:^\times V_j=\left[O,V_j\right],\label{Eq:define-hyper-cross}
\\
O\:^\circ V_j=\left\{O,V_j\right\},\label{Eq:define-hyper-circ}
\\
O\:^C V_j=OV_j.\label{Eq:define-hyper-C}
\end{eqnarray}
\end{subequations}

\vspace*{\fill}
\pagebreak

\section{Derivation of the Redfield Equation Comprising Photoexcitation}

We start from the central result of our analysis
\begin{equation}
\label{Eq:final-rho-ee-formal}
\rho_{ee}^{(I)}(t)=\int_{t_0}^t\dif\tau_2\int_{t_0}^{\tau_2}\dif\tau_1\:
 \overrightarrow{\mathcal{U}}_\mathrm{red}^{(I)}(t,\tau_2,\tau_1)A^{(I)}(\tau_2,\tau_1)+\mathrm{H.c.}
\end{equation}
and specialize to the case of excitation by a weak laser pulse. Taking time derivative of Eq.~\eqref{Eq:final-rho-ee-formal}, we obtain
\begin{equation}
\label{Eq:dif-t-rho-ee-general}
\partial_t\rho_{ee}^{(I)}(t)=-\frac{\im}{\hbar}\rho^{(I)}_{eg}(t)\boldsymbol{\mu}_{ge}^{(I)}(t)\cdot\boldsymbol{\mathcal{E}}^{(-)}(t)
+\int_{t_0}^t\dif\tau_2\int_{t_0}^{\tau_2}\dif\tau_1\:\partial_t\overrightarrow{\mathcal{U}}_\mathrm{red}^{(I)}(t,\tau_2,\tau_1)A^{(I)}(\tau_2,\tau_1)+\mathrm{H.c.}
\end{equation}
The time derivative of the reduced evolution superoperator reads as
\begin{equation}
\label{Eq:partial-t-right-U-towards-redfield}
\begin{split}
&\partial_t\overrightarrow{\mathcal{U}}_\mathrm{red}^{(I)}(t,\tau_2,\tau_1)=-\sum_j V_j^{(I)}(t)^\times\\
&\times T\left\{
\left[
\int_0^{t-\tau_2}\dif s\left(\frac{C_j^r(s)}{\hbar^2}V_j^{(I)}(t-s)^\times+\im\frac{C_j^i(s)}{\hbar^2}V_j^{(I)}(t-s)^\circ\right)+
\int_{t-\tau_2}^{t-\tau_1}\dif s\:\frac{C_j(s)}{\hbar^2}V_j^{(I)}(t-s)^C
\right]\overrightarrow{\mathcal{U}}_\mathrm{red}^{(I)}(t,\tau_2,\tau_1)
\right\}
\end{split}
\end{equation}

If we now assume that the characteristic decay time of the bath correlation function $C_j(t)$ is short compared to the time scales of the dynamics we are interested in, we can formally set $t-\tau_2\to+\infty$. Then, the second integral on the right-hand side of Eq.~\eqref{Eq:partial-t-right-U-towards-redfield} is equal to zero, while in the first integral we can invoke Markovian approximation,~\cite{May-Kuhn-book} which enables us to formally move hyperoperators $V_j^{(I)}(t-s)^{\times/\circ}$ in front of the $T$ sign. Transferring back to the Schr\"{o}dinger picture, we obtain
\begin{equation}
% \begin{split}
\partial_t\rho_{ee}(t)=-\frac{\im}{\hbar}\left[H_M,\rho_{ee}(t)\right]-\frac{\im}{\hbar}\rho_{eg}(t)\boldsymbol{\mu}_{ge}\cdot\boldsymbol{\mathcal{E}}^{(-)}(t)+\frac{\im}{\hbar}\boldsymbol{\mathcal{E}}^{(+)}(t)\cdot\boldsymbol{\mu}_{eg}\rho_{eg}^\dagger(t)-\sum_j V_j^\times\left[\Lambda_j\rho_{ee}(t)-\rho_{ee}(t)\Lambda_j^\dagger\right],
% \end{split}
\end{equation}
where
\begin{equation}
\Lambda_j=\int_0^{+\infty}\dif s\:\frac{C_j(s)}{\hbar^2}\:
\e^{-\im H_M s/\hbar}V_j\:\e^{\im H_Ms/\hbar}.
\end{equation}
In an analogous manner, we obtain the following second-order equation for optical coherences
\begin{equation}
% \begin{split}
\partial_t\rho_{eg}(t)=-\frac{\im}{\hbar}\left[H_M,\rho_{eg}(t)\right]+\frac{\im}{\hbar}\boldsymbol{\mu}_{eg}\cdot\boldsymbol{\mathcal{E}}^{(+)}(t)|g\rangle\langle g|-\sum_j V_j\Lambda_j\rho_{eg}(t).
% \end{split}
\end{equation}

Further manipulations towards the Redfield equation take place in the excitonic basis $\{|x\rangle\}$. These are quite standard,~\cite{May-Kuhn-book} and result in Eqs.~(37) and~(38) of the main text.

\newpage

\section{Excitation by Weak Incoherent Light: Level of Quantum Optical Master Equations}
Here, we demonstrate in more detail how the excitation by weak incoherent light should be handled on the level of quantum optical master equations to produce the HEOM.
We start from the general expression for $\rho^{(I)}_{ee}(t)$ [Eq.~\eqref{Eq:final-rho-ee-formal}] in which we insert the following the radiation correlation function~\cite{Mandel-Wolf-book}
\begin{equation}
\label{Eq:G-1-ij-thermal}
 G^{(1)}_{ij}(\tau_2,\tau_1)=\delta_{ij}\:\frac{\hbar}{6\pi^2\varepsilon_0 c^3}\int_0^{+\infty}\dif\omega\:\frac{\omega^3}{\e^{\beta_R\hbar\omega}-1}\:\e^{\im\omega(\tau_2-\tau_1)}
\end{equation}
to obtain
\begin{equation}
\begin{split}
\rho^{(I)}_{ee}(t)&=\int_{t_0}^t\dif\tau_2\int_{t_0}^{\tau_2}\dif\tau_1\:\overrightarrow{\mathcal{U}}^{(I)}_\mathrm{red}(t,\tau_2,\tau_1)\frac{1}{\hbar^2}\left[\boldsymbol{\mu}_{eg}^{(I)}(\tau_1)\cdot\boldsymbol{\mu}_{ge}^{(I)}(\tau_2)\right]\frac{\hbar}{6\pi^2\varepsilon_0c^3}\int_0^{+\infty}\dif\omega\:\omega^3\:n_\mathrm{BE}(\omega,T_R)\:\e^{\im\omega(\tau_2-\tau_1)}+\\
&+\text{H.c.}
\end{split}
\end{equation}
The radiation correlation function in Eq.~\eqref{Eq:G-1-ij-thermal} is computed for the three-dimensional photon gas at temperature $T_R=(k_B\beta_R)^{-1}$, and $n_\mathrm{BE}(\omega_{x},T_R)=\left(\e^{\beta_R\hbar\omega_x}-1\right)^{-1}$.

As is usual when the interaction of matter with electromagnetic radiation is treated on the quantum optical level,~\cite{Breuer-Petruccione-book} further developments should be conducted in the eigenbasis of $H_M$, i.e., in the excitonic basis $\{|x\rangle\}$, whose basis vectors satisfy $H_M|x\rangle=\hbar\omega_x|x\rangle$. Introducing time intervals $t_1=\tau_2-\tau_1$ and $t_2=t-\tau_2$, and transferring to the excitonic basis, we obtain
\begin{equation}
\label{Eq:quant-opt-inter}
\begin{split}
\rho^{(I)}_{ee}(t)&=\frac{1}{\hbar^2}\sum_{\bar x x}\left(\boldsymbol{\mu}_{\bar x}\cdot\boldsymbol{\mu}_x^*\right)\int_0^{t-t_0}\dif t_2\:\e^{\im\omega_{\bar x}(t-t_2-t_0)}\:\e^{-\im\omega_x(t-t_2-t_0)}\int_0^{+\infty}\dif\omega\:\frac{\hbar}{6\pi^2\varepsilon_0 c^3}\:\omega^3\:n_\mathrm{BE}(\omega,T_R)\times\\
&\times\int_0^{t-t_0}\dif t_1\:\e^{\im(\omega-\omega_{\bar x})t_1}
\overrightarrow{\mathcal{U}}^{(I)}_\mathrm{red}(t,t-t_2,t-t_2-t_1)|\bar x\rangle\langle x|+\\
&+\text{H.c.}
\end{split}
\end{equation}
Further steps are inspired by the Weisskopf--Wigner approximation.~\cite{Scully-Zubairy-book} Namely, the integral over $t_1$ contains the phase factor $\e^{\im(\omega-\omega_{\bar x})t_1}$ that exhibits oscillatory behaviour unless $\omega\approx\omega_{\bar x}$. By employing this approximation in Eq.~\eqref{Eq:quant-opt-inter}, the integral over $\omega$ reduces to
\begin{equation}
\int_0^{+\infty}\dif\omega\:\e^{\im\omega t_1}=\pi\delta(t_1)+\im\mathcal{P}\left(\frac{1}{t_1}\right),
\end{equation}
where $\mathcal{P}$ denotes the Cauchy principal value. In the spirit of Weisskopf--Wigner approximation, the part containing the principal-value sign is neglected, and the integration over $t_1$ is performed to arrive at
\begin{equation}
\begin{split}
\rho^{(I)}_{ee}(t)&=\sum_{\bar x x}\left(\boldsymbol{\mu}_{\bar x}\cdot\boldsymbol{\mu}_x^*\right)\int_{t_0}^t\dif\tau_2\:\frac{\omega_{\bar x}^3}{6\pi\varepsilon_0\hbar c^3}\:n_\mathrm{BE}(\omega,T_R)\overrightarrow{\mathcal{U}}^{(I)}_\mathrm{red}(t,\tau_2,\tau_2)\:\e^{\im\omega_{\bar x}(\tau_2-t_0)}|\bar x\rangle\langle x|\:\e^{-\im\omega_x(\tau_2-t_0)}+\\
&+\text{H.c.}
\end{split}
\end{equation}
where we have also restored the interaction instant $\tau_2$ as the integration variable. It is now apparent that the quantum-optical limit is intimately connected to the white-noise model considered in the main text. In both cases, the coherence time of the radiation is assumed to be negligible compared to other relevant time scales in the problem, so that both interactions with the radiation take place at the same instant $\tau_2$.

The Hermitian conjugate of the first summand in the last equation is easily calculated by noting that the exact reduced propagator $\overrightarrow{\mathcal{U}}^{(I)}_\mathrm{red}(t,\tau_2,\tau_2)$, which actually propagates only the excited-state sector of the reduced density matrix, satisfies
\begin{equation}
A\overleftarrow{\mathcal{U}}^{(I)}_\mathrm{red}(t,\tau_2,\tau_2)=\overrightarrow{\mathcal{U}}^{(I)}_\mathrm{red}(t,\tau_2,\tau_2)A,
\end{equation}
for arbitrary purely electronic operator $A$. Introducing the spontaneous emission rate $\Gamma_x$ from excitonic state $|x\rangle$
\begin{equation}
\Gamma_x=\frac{1}{4\pi\varepsilon_0}\:\frac{4\omega_x^3\left|\boldsymbol{\mu}_x\right|^2}{3\hbar c^3},
\end{equation}
the final result for the excited-state sector of the RDM in the interaction picture reads as
\begin{equation}
\label{Eq:qome-general}
\begin{split}
\rho^{(I)}_{ee}(t)&=\int_{t_0}^t\dif\tau_2\:
\overrightarrow{\mathcal{U}}^{(I)}_\mathrm{red}(t,\tau_2,\tau_2)\times\\
&\times\sum_{\bar x x}
\left[\frac{\boldsymbol{\mu}_{\bar x}\cdot\boldsymbol{\mu}_x^*}{\left|\boldsymbol{\mu}_{\bar x}\right|^2}\frac{1}{2}\Gamma_{\bar x}n_\mathrm{BE}(\omega_{\bar x},T_R)+\frac{\boldsymbol{\mu}_{\bar x}\cdot\boldsymbol{\mu}_x^*}{\left|\boldsymbol{\mu}_{x}\right|^2}\frac{1}{2}\Gamma_{x}n_\mathrm{BE}(\omega_{x},T_R)\right]\e^{\im\omega_{\bar x}(\tau_2-t_0)}|\bar x\rangle\langle x|\:\e^{-\im\omega_x(\tau_2-t_0)}.
\end{split}
\end{equation}
One can now formulate in the usual manner the HEOM that replaces Eq.~\eqref{Eq:qome-general}. In doing so, we immediately realize that only the equation of motion for RDM has the source term describing the generation of excitations from the ground state, while ADMs do not possess such a term. In greater detail, the interaction-picture ADM labeled by vector $\mathbf{n}$ assumes the form
\begin{equation}
\begin{split}
\sigma_{ee,\mathbf{n}}^{(I)}(t)&=\int_{t_0}^t\dif\tau_2\:T\left\{\prod_j\prod_m\left[\int_{\tau_2}^t\dif s\:\e^{-\mu_{j,m}(t-s)}\left(\im\frac{c_{j,m}^r}{\hbar^2}V_j^{(I)}(s)^\times-\frac{c_{j,m}^i}{\hbar^2}V_j^{(I)}(s)^\circ\right)\right]^{n_{j,m}}\overrightarrow{\mathcal{U}}^{(I)}_\mathrm{red}(t,\tau_2,\tau_2)\right\}\times\\
&\times\sum_{\bar x x}
\left[\frac{\boldsymbol{\mu}_{\bar x}\cdot\boldsymbol{\mu}_x^*}{\left|\boldsymbol{\mu}_{\bar x}\right|^2}\frac{1}{2}\Gamma_{\bar x}n_\mathrm{BE}(\omega_{\bar x},T_R)+\frac{\boldsymbol{\mu}_{\bar x}\cdot\boldsymbol{\mu}_x^*}{\left|\boldsymbol{\mu}_{x}\right|^2}\frac{1}{2}\Gamma_{x}n_\mathrm{BE}(\omega_{x},T_R)\right]\e^{\im\omega_{\bar x}(\tau_2-t_0)}|\bar x\rangle\langle x|\:\e^{-\im\omega_x(\tau_2-t_0)}
\end{split}
\end{equation}
while its equation of motion reads as
\begin{equation}
\label{Eq:n-barx-x-quantum-optical}
\begin{split}
&\partial_t\sigma_{ee,\mathbf{n}}(t)=-\frac{\im}{\hbar}\left[H_M,\sigma_{ee,\mathbf{n}}(t)\right]-\left(\sum_j\sum_m n_{j,m}\mu_{j,m}\right)\sigma_{ee,\mathbf{n}}(t)\\
  &+\delta_{\mathbf{n},\mathbf{0}}\sum_{\bar x x}\frac{\boldsymbol{\mu}_{\bar x}\cdot\boldsymbol{\mu}_x^*}{\left|\boldsymbol{\mu}_{\bar x}\right|^2}\frac{1}{2}\Gamma_{\bar x}n_\mathrm{BE}(\omega_{\bar x},T_R)|\bar x\rangle\langle x|+\delta_{\mathbf{n},\mathbf{0}}\sum_{\bar x x}\frac{\boldsymbol{\mu}_{\bar x}\cdot\boldsymbol{\mu}_x^*}{\left|\boldsymbol{\mu}_{x}\right|^2}\frac{1}{2}\Gamma_{x}n_\mathrm{BE}(\omega_{x},T_R)|\bar x\rangle\langle x|\\
  &+\im\sum_j\sum_m\left[V_j,\sigma_{ee,\mathbf{n}_{j,m}^+}(t)\right]+\im\sum_j\sum_m n_{j,m}\left(\frac{c_{j,m}}{\hbar^2}\:V_j\sigma_{ee,\mathbf{n}_{j,m}^-}(t)-
  \frac{c_{j,m}^*}{\hbar^2}\sigma_{ee,\mathbf{n}_{j,m}^-}(t)V_j\right).
\end{split}
\end{equation}

In Eq.~\eqref{Eq:n-barx-x-quantum-optical}, the generation of excited-state populations and intraband coherences from the ground state is described by the source terms containing excitonic dipole moments, spontaneous emission rates, and photon occupation numbers. The rate at which the population of excitonic state $|x\rangle$ is generated from the ground state assumes the familiar form
\begin{equation}
\label{Eq:pop-gen-rate-quantum-optical}
\left(\partial_t n_{xx}(t)\right)_\mathrm{source}=\Gamma_x n_\mathrm{BE}(\omega_x,T_R),
\end{equation}
where the spontaneous emission rate $\Gamma_x$ is multiplied by the Bose--Einsten factor, which is characteristic for the absorption of one photon of energy $\hbar\omega_x$. The rate at which the intraband coherence between excitonic states $|\bar x\rangle$ and $|x\rangle$ ($\bar x\neq x$) is generated from the ground state contains factors $\boldsymbol{\mu}_{\bar x}\cdot\boldsymbol{\mu}_x^*$ describing the alignment of the corresponding transition dipole moments
\begin{equation}
% \begin{split}
\left(\partial_t n_{\bar x x}(t)\right)_\mathrm{source}=\frac{\boldsymbol{\mu}_{\bar x}\cdot\boldsymbol{\mu}_x^*}{\left|\boldsymbol{\mu}_{\bar x}\right|^2}\frac{1}{2}\Gamma_{\bar x}n_\mathrm{BE}(\omega_{\bar x},T_R)+\frac{\boldsymbol{\mu}_{\bar x}\cdot\boldsymbol{\mu}_x^*}{\left|\boldsymbol{\mu}_{x}\right|^2}\frac{1}{2}\Gamma_{x}n_\mathrm{BE}(\omega_{x},T_R).
% \end{split}
\end{equation}
Interestingly, these source terms are present exclusively in the equation for the RDM, as indicated by the presence of the Kronecker delta $\delta_{\mathbf{n},\mathbf{0}}$. At first sight, this is very different from the description of the light--matter interaction on the quantum-optical level in, e.g., Ref.~\onlinecite{JPhysB.51.054002}, which is inspired by the combined Born--Markov--HEOM approach developed in Ref.~\onlinecite{JChemTheoryComput.7.2166}. There, each level of the hierarchy contains source terms similar to the ones we encounter in Eq.~\eqref{Eq:n-barx-x-quantum-optical} on the level of the RDM. Moreover, the quantum-optical source terms in Ref.~\onlinecite{JPhysB.51.054002} also feature the radiative recombination terms, which deplete excited-state populations and increase the ground-state population.

The reason for such differences lies in the fact that our treatment of the photoexcitation process starts from the unexcited system and is consistently up to the second order in the exciting field. Within our approximations, the ground-state population is always close to 1, while the excited-state populations are at least quadratic in the weak exciting field and are much smaller than 1, compare to the scaling laws under semiclassical light--matter interactions [Eqs.~(13)] presented in the main text. We have already used similar arguments in the main text to transform the HEOM that does not take into account scaling laws to the HEOM that is consistently up to the second order in the optical field. Here, on the quantum-optical level, the generation rate of excited-state populations, Eq.~\eqref{Eq:pop-gen-rate-quantum-optical}, implicitly contains our assumption that, at all times, the ground-state population differs from 1 by a quantity that is at least quadratic in the exciting field. Similar terms for higher-tier ADOs are absent in our treatment simply because their ground-state expectation values are approximately 0 at all times. In a similar vein, our treatment cannot capture radiative recombination from excited states because that process is at least of the fourth order in the field.

\newpage
\bibliography{apssamp}% Produces the bibliography via BibTeX.